\documentclass[pre,aps,showpacs,nofootinbib]{revtex4}
\usepackage{graphicx}
\usepackage{natbib}
\usepackage{bm}
\usepackage{amssymb}
\usepackage{amsmath}
\usepackage{euscript}
\usepackage{esint}
\usepackage{color}
\usepackage{ulem}

\begin{document}

\title{On-Shell Description of Unsteady Flames}

\author{Guy Joulin$^1$, Hazem El-Rabii$^1$ and
Kirill~A.~Kazakov$^2$\thanks{E-mail: $Kirill@theor.phys.msu.su$}}%
\affiliation{%
$^1$Laboratoire de Combustion et de D\'etonique, CNRS/ENSMA, 1 av. Cl\'ement Ader, 86961 Futuroscope, Poitiers, France}%

\affiliation{$^2$Department of Theoretical Physics, Physics Faculty,
Moscow State University, 119899, Moscow, Russian Federation}

\begin{abstract}
The problem of non-perturbative description of unsteady premixed
flames with arbitrary gas expansion is solved in the two-dimensional
case. Considering the flame as a surface of discontinuity with
arbitrary local burning rate and gas velocity jumps given on it, we
show that the front dynamics can be determined without having to
solve the flow equations in the bulk. On the basis of the Thomson
circulation theorem, an implicit integral representation of the gas
velocity downstream is constructed. It is then simplified by a
successive stripping of the potential contributions to obtain an
explicit expression for the vortex component near the flame front.
We prove that the unknown potential component is left bounded and
divergence-free by this procedure, and hence can be eliminated using
the dispersion relation for its on-shell value (i.e., the value
along the flame front). The resulting system of integro-differential
equations relates the on-shell fuel velocity and the front position.
As limiting cases, these equations contain all theoretical results
on flame dynamics established so far, including the linear equation
describing the Darrieus-Landau instability of planar flames, and the
nonlinear Sivashinsky-Clavin equation for flames with weak gas
expansion.
\end{abstract}
\pacs{47.20.-k, 47.32.-y, 82.33.Vx} \maketitle


\unitlength=1pt

\noindent

\section{Introduction}

Although the elementary physical mechanisms underlying flame
propagation are well-understood by now, a global mathematical
description of the process is extremely difficult. The reasons --
which flames share with condensation discontinuities in
supersaturated vapors \citep{landau1}, ablation fronts driven by
lasers or X-rays \citep{clavin}, thermonuclear waves in type-Ia
supernov\ae~\citep{hillebrandt} or rapid decomposition of explosive
liquids \citep{landau1} -- can be summarized briefly: all involve
propagating free-boundaries crossed by nonzero mass fluxes and
separating subsonic flows of markedly different densities. In view
of this it is not surprising that, since their identification as
self-propagating slow deflagrations (as opposed to detonations) in
the $1880$s \citep{mallard}, flames had to wait some fifty years
before their theoretical study began at the simplest level of linear
stability analysis of planar flames \citep{darrieus,landau}. It took
some forty years more for the first consistent account of nonlinear
effects to appear. It was shown by \cite{siv1} how these effects can
be described in the case when the fresh to burnt gas density ratio,
$\theta,$ is close to unity. The latter condition is a principle
limitation for a {\it perturbative} treatment of nonlinear
saturation phenomena. Indeed, the weak nonlinearity expansions are
simply self-contradictory in the case of steady flames with a finite
$(\theta-1)$ \citep{kazakov3}. Despite the fact that the practically
relevant values of $\theta$ are $5$ to $8,$ the small $(\theta - 1)$
approximation has so far been the sole theoretical
method\footnote{We do not consider models inconsistent from the
standpoint of the fundamental equations, such as for instance the
Frankel's \citep{frankel1990} potential-flow model. The latter
assumes finite values of $(\theta-1)$ and zero vorticity production
in the flame front at the same time; these two assumptions are only
consistent in the lowest order of the small $(\theta-1)$ expansion,
in which case the Frankel equation reduces to the Sivashinsky
equation \citep{siv1}.} available to handle the flame front
dynamics.

One of the essential difficulties encountered in any analytic
treatment of flames is the virtual impossibility to solve the flow
equations governing the dynamics of the burned gases. This is an
exceedingly complicated problem which requires finding solutions to
a system of nonlinear partial differential equations in the regions
ahead of the flame front and behind it, to be chosen so as to
satisfy a number of jump conditions expressing the conservation of
mass, energy and momentum across the moving front. The front
dynamics itself is determined by the so-called evolution equation
describing the local fuel consumption rate as a functional of the
fuel velocity distribution along the front and the front shape
\citep{markstein1951}. Even if the gas flow is potential upstream,
as is the case for flames propagating in an initially quiescent
fluid, this property is lost in the downstream region because
vorticity is generated by the curved flame front, so that the
problem of solving the flow equations is faced in its full
generality.

Struggling with this problem is indeed unavoidable if one is
interested in the explicit structure of the burnt gas flow. However,
it is the evolution of the flame front, its position and shape that
usually constitute the main concern in practice. This limitation of
the problem raises naturally the following {\it dilemma}
\citep[see][]{kazakov1,kazakov2}: On the one hand, deflagration is
an essentially non-local and nonlinear process with all its
complications mentioned above; on the other hand, this non-locality
itself is determined by the flame front configuration and the gas
velocity distribution along it, which play the role of boundary
conditions for the flow equations and thus control the bulk flow. In
such circumstances, is it really necessary to know explicitly the
flow structure in the entire downstream region in order to describe
the front evolution in a closed form, i.e. in a form involving only
quantities defined on the flame front?

As to the steady configurations, this question was settled out in
the negative by \cite{kazakov1,kazakov2}. More precisely, it was
shown that the only piece of information about the gas flow
downstream that is really necessary to derive an equation for the
flame front position is the value of the vortex component of burnt
gas along the front or, using the terminology used by
\cite{kazakov2}, the {\it on-shell value} of this component. The
remaining unknown potential component of the gas velocity is
excluded eventually using a ``dispersion relation'' for its on-shell
values (expressing analyticity of this component in the downstream
region), thus providing one with a description of flame shapes in a
form which is closed in the above-mentioned sense. The purpose of
the present paper is to generalize this construction to the case of
unsteady flame propagation. Although the procedure is essentially
the same as in the steady case, a subtle point is worth to be
emphasized. According to \cite{kazakov1,kazakov2}, the derivative of
the vortex component along the flame front is local, i.e., its value
at a given point is a function of the on-shell fuel velocity, its
derivatives, and the front shape at the same point. In view of this,
one might expect the generalization to the unsteady case to be
purely kinematic, namely, that it would amount to rewriting the
steady equation in terms of the gas velocities relative to the local
front velocity. We will see, however, that this is not so because of
a peculiar -- yet unavoidable -- spatial non-locality of the vortex
component, which appears naturally in the unsteady case to account
for the effect of the flame history on its current evolution.
Because of this complication, one has to be more careful with the
spatial integrations involved in the definition of the vorticity
mode. To define the improper integrals along the flame front, we use
an intermediate regularization. Specifically, we introduce an
exponential damping of the contributions coming from remote parts of
the front. This regularization is eventually removed using analytic
continuation to the case of zero damping.

The paper is organized as follows. We first construct an implicit
integral of the flow equations downstream on the basis of the
Thomson circulation theorem, which expresses the gas velocity in
terms of its boundary values and vorticity distribution behind the
front. This is done in Sec.~\ref{flowequations}. The integral
representation is then used in Sec.~\ref{structure} to obtain an
expression for the vortex mode of the gas velocity near the front,
which is accomplished by successive stripping burnt gas velocity of
potential contributions. We prove that the unknown potential
component is left bounded and divergence-free by this procedure.
Hence, it can be eliminated using the dispersion relation for its
on-shell value, thus leading to the main integro-differential
equation written down in Sec.~\ref{closed}. This equation relates
the on-shell value of the fuel velocity and the flame front
position, and together with the evolution equation constitutes the
closed system for these quantities. Finally, it is verified in
Sec.~\ref{verification} that the derived equation contains as simple
limiting cases all known theoretical results on flame dynamics,
namely, the linear equation describing Darrieus-Landau instability
of planar flames \citep{darrieus,landau}, including the case when
the flame propagates in a time-dependent gravitational field
\citep{acoust}, the nonlinear Sivashinsky-Clavin equation for flames
with weak gas expansion \citep{siv1,sivclav}, and the stationary
equation derived by \cite{kazakov2}. Appendix contains a consistency
check for the results obtained in Sec.~\ref{structure}.

\section{Integral representation of the flow equations}\label{flowequations}

Consider a flame propagating in an initially uniform premixed ideal
fluid. Our analysis below relies substantially on the well-known
Thomson theorem stating that circulation of the gas velocity over
any closed material contour drawn in an ideal fluid is conserved as
it is convected. This statement takes an especially simple form in
the case of two-dimensional ($2D$) incompressible flows, since not
only the circulation itself, but also the value of vorticity carried
by any fluid element is then conserved. Since the space
dimensionality is not that important in the {\it formulation} (not
the resolution) of the dilemma mentioned in the Introduction, we
will be concerned in what follows with the simpler $2D$ case. We
will further specify our analysis to flames propagating in a channel
of constant width $b.$ Let the Cartesian coordinates $(x,y)$ be
chosen so that the $y$-axis is parallel to the tube walls, $y = -
\infty$ being in the fresh fuel. These coordinates will be measured
in units of the channel width, while the fluid velocity, $\bm{v} =
(v_1,v_2),$ in units of the velocity of a plane flame front relative
to the fuel. It will be sometimes useful to denote the Cartesian
components of $\bm{v}$ by $(w,u).$ Finally, the fluid density will
be normalized on the fuel density, $\theta> 1$ denoting its ratio to
that of the burnt gas.

It will be more appropriate for our purposes to reformulate the
problem under consideration as a problem of propagation of an
unbounded spatially-periodic flame. Namely, assuming the channel
walls ideal, given a flame configuration described by the functions
$f(x,t), \bm{v}(x,y,t),$ $x \in [0,+1],$ where $f(x,t)$ denotes the
flame front position at time instant $t,$ using the boundary
conditions $f'=0, w=0$ for $x = 0,1,$ we continue this configuration
to the domain $x \in [-1,0]$ according to
\begin{eqnarray}\label{antisym}
f(x,t) = f(-x,t)\,, \qquad w(x,y,t) = - w(-x,y,t)\,, \qquad u(x,y,t)
= u(-x,y,t)\,,
\end{eqnarray}
\noindent and then periodically continue it to the whole $x$-axis.
Note that having imposed the boundary condition $f'(0) = f'(1) = 0,$
we thereby exclude the possibility of stagnation zone formation near
the end points of the flame front (see \cite{zel1} for detail). We
also assume that the flame is stable with respect to the short
wavelength perturbations i.e., that there is a short wavelength
cutoff, $\lambda_c.$ This cutoff ensures smoothness of the functions
under consideration. In particular, it prevents the development of
singularities of the front shape such as the edge points which would
occur otherwise \citep{zel1}, leading to discontinuities in the
values of the flow variables or their derivatives. That $\lambda_c$
often exceeds the actual thickness of the flame preheat zone
significantly \citep{searby1991} has yet another virtue: the
Reynolds number based on $\lambda_c$ and the fuel properties is
typically over $\sim 10^2$, and hence is fairly large when based
upon the width ($>\lambda_c$ or $\gg \lambda_c$) of the channel
where the flame studied below is meant to propagate. It then makes
sense to model the flame as a surface (or line in 2-D) equipped with
a local $\lambda_c$-dependent propagation law, and embedded in ideal
fluid flows. We shall return to this issue in the final section of
the paper, merely mentioning here that viscosity effects are known
from direct numerical simulations \citep{bychkov2000} to have
negligible influence on the shape and the speed of steady curved
flames.

In our formulation, the flow velocity obeys the following equations
in the bulk
\begin{eqnarray}\label{flow1}
\frac{\partial v_i}{\partial x_i} &=& 0\,,
\\ \frac{\partial\sigma}{\partial t}
+ v_i\frac{\partial \sigma}{\partial x_i} &=& 0 \,, \label{flow2}
\end{eqnarray}
\noindent where $(x_1,x_2) = (x,y),$
\begin{eqnarray}\label{flow3}
\sigma = \frac{\partial v_2}{\partial x} - \frac{\partial
v_1}{\partial y}
\end{eqnarray}
\noindent is the vorticity, and summation over repeated indices is
understood. The first equation expresses continuity of the
incompressible flows, while the other the Thomson theorem. It is not
difficult to see that $\bm{v}$ satisfying Eqs.~(\ref{flow1}),
(\ref{flow3}) can be written in the following integral form (Cf.
derivation of Eq.~(9) in \citealt{kazakov2})
\begin{eqnarray}\label{vint}
v_i &=& \varepsilon_{ik}\partial_k \int\limits_{\Lambda}d
l_l~\varepsilon_{lm} v_m\frac{\ln r}{2\pi} -
\partial_i\int\limits_{\Lambda}d l_k~v_k\frac{\ln r}{2\pi} -
\varepsilon_{ik}\partial_k \int\limits_{\Sigma}d s~\frac{\ln
r}{2\pi} \sigma \,,
\end{eqnarray}
\noindent where $\varepsilon_{ik} = - \varepsilon_{ki},\
\varepsilon_{12} = + 1,$ $\partial_i = \partial/\partial x_i.$
$\Sigma$ and $\Lambda$ denote any part of the downstream region and
its boundary, respectively. $r$ is the distance between an
infinitesimal fluid element $ds$ at the point
$(\tilde{x},\tilde{y})$ and the point of observation
$\bm{x}=(x,y)\in \Sigma,$ $r^2 = (x_i - \tilde{x}_i)^2,$ and $d l_i$
is the line element normal to $\Lambda$ and directed outwards of
$\Sigma.$ Indeed, to evaluate divergence of the right hand side of
Eq.~(\ref{vint}), we use the relations
$\partial_i\varepsilon_{ik}\partial_k \equiv 0,$
\begin{eqnarray}\label{green}
\partial^2_{k}\ln r =
2\pi\delta(x-\tilde{x})\delta(y-\tilde{y})\,,
\end{eqnarray}
\noindent where $\delta(x)$ is the Dirac delta-function. It follows
that $\partial_i v_i = 0$ for any point $\bm{x}$ inside $\Sigma.$
Similarly, evaluating the curl of Eq.~(\ref{vint}) with the help of
Eq.~(\ref{green}) and $\varepsilon_{ik} \varepsilon_{im} =
\delta_{km}$ gives the identity $\varepsilon_{ik}\partial_iv_k =
\sigma.$ We are now to employ Eq.~(\ref{flow2}) to rewrite the last
term in Eq.~(\ref{vint}) as an integral over fluid particle
trajectories.

\begin{figure}
\centering
\includegraphics[width=.6\textwidth]{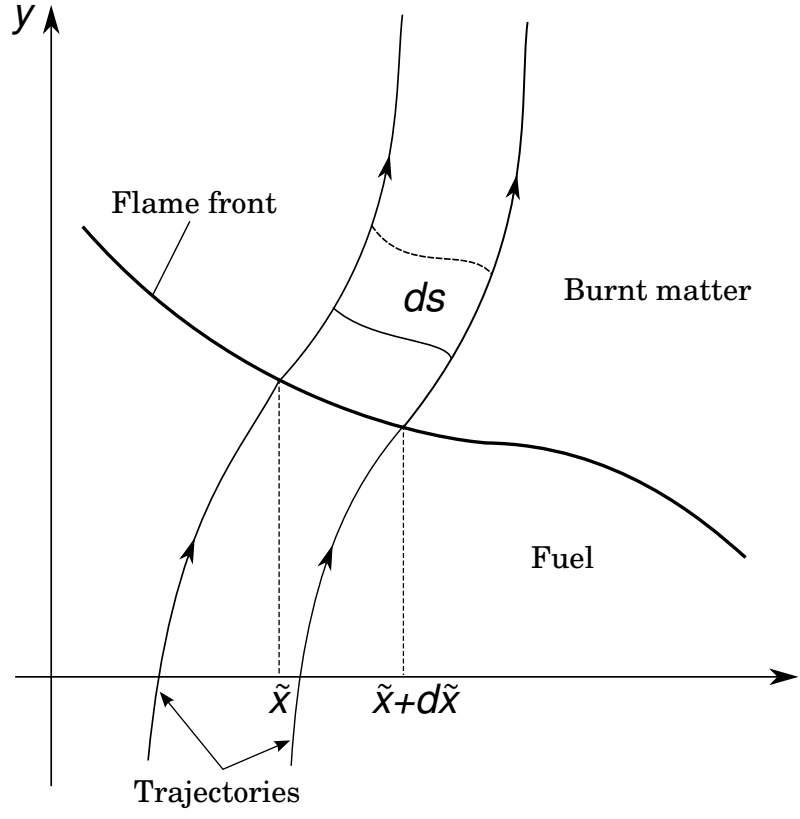}
\caption{Elementary decomposition of the flow downstream used in the
derivation of the expression (\ref{vint1}).}\label{fig1}
\end{figure}

To this end, we specify that $\Sigma$ is spanned by fluid elements
that crossed the flame front between two points on it with fixed
abscissas $\tilde{x}_1 = - A$ and $\tilde{x}_2 = + A$ during the
time interval $[-T,t],$ where $t$ is the given time instant the flow
is observed at (see Fig.~\ref{fig1}). Here $A,T$ are some positive
parameters tending eventually to infinity, so that $\Sigma$ will
then fill the whole downstream region: $\Sigma =
\{\tilde{x},\tilde{y}: \tilde{y}
> f(\tilde{x},t)\}.$ But for the time being, we
keep them finite. The improper integral over an infinite $\Sigma$
will be defined later, in Sec.~\ref{vortdefinition}. Then we have,
by virtue of the vorticity conservation,
\begin{eqnarray}\label{vint1}
\lefteqn{\int\limits_{\Sigma}\!d s~\sigma\ln r =}\nonumber\\
& & \int\limits_{-A}^{+A}\!\!d\tilde{x}\int\limits_{-T}^{t}\!\!d\tau
N(\tilde{x},\tau)\bar{v}^n_+(\tilde{x},\tau)\sigma_+(\tilde{x},\tau)
\ln\left\{[x - X(\tilde{x},t,\tau)]^2 + [y -
Y(\tilde{x},t,\tau)]^2\right\}^{1/2}\,,
\end{eqnarray}
\noindent where $N = \sqrt{1 + \left(\partial f/ \partial
x\right)^2}\,,$ $\bar{v}^n_+ = \bar{v}_{i+} n_i$ is the normal burnt
gas velocity relative to the flame front,
$$\bar{\bm{v}}_+ =
(w_+,\bar{u}_+)\,, \quad \bar{u}_+(x,t) \equiv u_+(x,t) -
\frac{\partial f(x,t)}{\partial t}\,,$$ $n_i$ is the unit vector
normal to the front (pointing to the burnt matter), and
$(X(\tilde{x},t,\tau),$ $Y(\tilde{x},t,\tau))$ is the current
position of a fluid element that crossed the point $(\tilde{x},
f(\tilde{x},\tau))$ on the flame front at $\tau.$ It is taken into
account in Eq.~(\ref{vint1}) that the ``volume'' $ds$ of this
element is conserved in view of the flow incompressibility, and
hence can be written as $d\tilde{x}d\tau
N(\tilde{x},\tau)\bar{v}^n_+(\tilde{x},\tau).$ Changing the
integration variable, $\tau \to t-\tau,$ in the expression
(\ref{vint1}), and substituting it into Eq.~(\ref{vint}) gives
\begin{eqnarray}\label{vint2}
v_i &=& \varepsilon_{ik}\partial_k \int\limits_{\Lambda}d
l_l~\varepsilon_{lm} v_m\frac{\ln r}{2\pi} -
\partial_i\int\limits_{\Lambda}d l_k~v_k\frac{\ln r}{2\pi} -
\frac{\varepsilon_{ik}}{2}\partial_k \int\limits_{-A}^{+A}d\tilde{x}
K(x,y,\tilde{x},t)\,,
\end{eqnarray}
\noindent where the integral kernel $K$ is defined by
\begin{eqnarray}\label{kernel}
K(x,y,\tilde{x},t) &=& \frac{1}{\pi}\int\limits_{0}^{T+t}\!\!d\tau
M(\tilde{x},t-\tau)\nonumber\\
&& \times\ln\left\{[x - X(\tilde{x},t,t-\tau)]^2 + [y -
Y(\tilde{x},t,t-\tau)]^2\right\}^{1/2}\,, \\
\nonumber\\M(\tilde{x},\tau) &\equiv&
N(\tilde{x},t)\bar{v}^n_+(\tilde{x},t)\sigma_+(\tilde{x},t)\,.
\label{mfunction}
\end{eqnarray}
\noindent

\section{Near-the-front structure of the vortex mode}\label{structure}

The integral representation (\ref{vint2}) of  the flow velocity
downstream is to be used below to obtain an expression for its
vortex component near the flame front. More precisely, we are going
to define a vortex component $v^v_i$ in a way that would allow an
explicit expression for its on-shell value in terms of the on-shell
gas velocity $v_{i+}$. For this purpose, we decompose the velocity
field as $$v_i = v^p_i + v^v_i\,, \quad i = 1,2,$$ where $v^p_i$ is
a potential component satisfying the following requirements
($D\bm{v}^p$ denotes any first-order spatial derivative of
$\bm{v}^p$):
\begin{itemize}
\item[a)]{${\rm div}D\bm{v}^p = 0,$}
\item[b)]{${\rm rot}D\bm{v}^p = 0,$}
\item[c)]{$D\bm{v}^p$ is bounded, in the sense that it
remains finite in the limit $A,T\to +\infty,$ i.e., for an
infinitely expanding $\Sigma$ region.}
\end{itemize} The on-shell expression for the
vorticity contribution will be obtained by a step-by-step
simplification of Eq.~(\ref{vint2}) throwing away potential fields
fulfilling the above conditions a)--c). Although this derivation
follows closely that of \cite{kazakov2}, we give it here in detail
to make clear the point where non-stationarity of the problem comes
into play.

Let the equality of two functions $\varphi_1(x,y),\, \varphi_2(x,y)$
up to a field satisfying a) -- c) be denoted by
$\varphi_1\stackrel{\circ}{=} \varphi_2.$ First of all, as we saw in
the preceding section, the first two terms on the right hand side of
Eq.~(\ref{vint2}) have vanishing curl and divergence, and hence
satisfy also the requirements a) and b). Furthermore, a simple power
counting shows that c) is also met. Indeed, consider part
$\Lambda\diagdown F$ of the contour $\Lambda,$ where $F$ denotes the
flame front. Representing this part as a semicircle with radius
$R\to \infty,$ we note that $D^2\ln r = O(1/R^2)$ for any given
$\bm{x}\in \Sigma$ ($D^2$ denotes any second spatial derivative).
Taking into account also that $\bm{v}=O(1),$ $dl = Rd\phi,$ where
$\phi\in (0,\pi)$ is the angular coordinate of the point
$\tilde{\bm{x}}$ on the semicircle, one sees that after spatial
differentiation, the two integrals over $\Lambda\diagdown F$ on the
right of Eq.~(\ref{vint2}) vanish in the limit $R\to \infty.$
Similar consideration shows that the same integrals over $F$ are
convergent, thus proving that $D\bm{v}^p$ remains bounded downstream
in the limit $A,T\to \infty.$ So we can write
\begin{eqnarray}\label{vint2v}
v_i \stackrel{\circ}{=} - \frac{\varepsilon_{ik}}{2}\partial_k
\int\limits_{-A}^{+A}d\tilde{x} K(x,y,\tilde{x},t)\,.
\end{eqnarray}
\noindent Next, we note that since we are eventually interested in
the on-shell value of the vortex component, we may take the
observation point $(x,y)$ as close to the flame front as we like,
i.e., $y\approx f(x,t),$ [$y>f(x,t)$]. The vortex component at such
points is determined by a contribution coming from the integration
over small $\tau$ and $\tilde{x} \approx x.$ Indeed, taking the curl
of Eq.~(\ref{vint2v}), and using Eq.~(\ref{green}) yields
\begin{eqnarray}
\varepsilon_{ik}\partial_iv_k  =
\int\limits_{-A}^{+A}d\tilde{x}\int\limits_{0}^{T+t}d\tau
M(\tilde{x},t-\tau) \delta(x - X[\tilde{x},t,t-\tau])\delta(y -
Y[\tilde{x},t,t-\tau])\,, \nonumber
\end{eqnarray}
\noindent which explicitly shows that a nonzero contribution to the
vorticity comes only from the point $\{\tilde{x},\tau\}$ obeying the
equations
\begin{eqnarray}\label{point1}
x - X[\tilde{x},t,t-\tau] = 0\,, \quad y - Y[\tilde{x},t,t-\tau] =
0\,,
\end{eqnarray}
\noindent which for $y\approx f(x,t)$ state that the point $(X,Y)$
is close to the flame front, and hence
\begin{eqnarray}
X(\tilde{x},t,t-\tau) & = & \tilde{x} + w_+(\tilde{x},t)\tau +
O(\tau^2),\nonumber\\ Y(\tilde{x},t,t-\tau) & = &
f(\tilde{x},t-\tau) + u_+(\tilde{x},t)\tau + O(\tau^2)\,.\nonumber
\end{eqnarray} Expanding also $f(\tilde{x},t-\tau)$ to the first order
in $\tau$ and omitting the symbols $O(\tau^2),$ we thus have the
following approximate expression for the fluid particle trajectory
\begin{eqnarray}\label{point2}&&
X(\tilde{x},t,t-\tau) \approx \tilde{x} + w_+(\tilde{x},t)\tau
\,,\quad Y(\tilde{x},t,t-\tau) \approx f(\tilde{x},t) +
\bar{u}_+(\tilde{x},t)\tau\,.
\end{eqnarray}
It follows that if these expressions are used instead of the exact
ones to calculate the kernel $K,$ we still have the true
distribution of vorticity near the flame front. Indeed, it was just
shown that any integration over values of $\{\tilde{x},\tau\}$ not
satisfying Eq.~(\ref{point1}), where Eq.~(\ref{point2}) is not valid
either, gives rise to a potential contribution. In particular, the
property b) of this contribution is preserved, and so is the
property a), since the right hand side of Eq.~(\ref{vint2v}) is
divergence-free identically whatever the form of the kernel.
Finally, it is not difficult to see that the condition c) is also
satisfied. Indeed, the above transformation of exact trajectories
into the straightened ones given by Eq.~(\ref{point2}) leaves the
expression (\ref{vint2v}) bounded. Therefore, the potential field
added in the course of this transformation is bounded, too. It
should be stressed that from now on we are concerned only with the
on-shell value of the vortex component, so the meaning of the symbol
$\stackrel{\circ}{=}$ should be further specified. Of course, the
transformation of trajectories changes the bulk vorticity
distribution, and thereby the velocity field downstream. However,
after the transformation as well as before it, integration over
finite $\tau$s in Eq.~(\ref{kernel}) results in a field that is
potential near the flame front. Both fields are of complicated
structure which is unknown in general, but since they are potential
near the front and bounded, we can use the on-shell value of their
difference to define a field satisfying a)--c) in the whole
downstream region. The existence of this field is guaranteed by the
Cauchy theorem. Namely, we use the Cauchy formula to construct the
field satisfying a)--c) as the analytic function with the given
boundary value.  It is in this sense that the above transformation
is said to respect the properties a)--c). In particular, the symbol
$\stackrel{\circ}{=}$ is used below to relate the on-shell values
(or near-the-front values, in the case of the integral kernel) of
functions that differ by a field satisfying a)--c) downstream.

We now proceed to an explicit evaluation of the integral kernel
(\ref{kernel}) which, after the transformations (\ref{point2}) are
performed, takes the form\footnote{When transforming the kernel
$K(x,y,\tilde{x},t),$ we use the same symbol $\stackrel{\circ}{=}$
to relate expressions that upon substitution in Eq.~(\ref{vint2v})
give rise to fields that are equal in the sense of
$\stackrel{\circ}{=}.$}
\begin{eqnarray}\label{kernel1}
K(x,y,\tilde{x},t) \stackrel{\circ}{=}
\frac{1}{\pi}\int\limits_{0}^{T+t}d\tau
M(\tilde{x},t-\tau)\ln\left\{ \bar{v}_+^2 \tau^2 - 2 (\bm{r}\cdot
\bar{\bm{v}}_+)\tau + r^2\right\}^{1/2}\,,
\end{eqnarray}
\noindent where $\bm{r} = (x-\tilde{x},y-f(\tilde{x},t)).$ The
integrand here involves $M(\tilde{x},t-\tau)$ which is an unknown
function of $\tau.$ In view of what has been said about the
near-the-front structure of the vorticity mode, one might think that
it would be sufficient to expand this function to the first order in
$\tau,$ and then calculate the integral. However, this operation is
not allowed as it would violate the condition c) and, as a result,
would yield erroneous predictions (see Sec. 5). In particular, the
on-shell value of the vorticity component cannot be found by setting
$\tau = 0$ in the argument of $M.$ There is no such problem in the
case of steadily propagating flames, as $M$ is then time-independent
in a frame of reference attached to the front. To overcome this
difficulty, we will explicitly extract the singular part of the
$\tau$-integral, which is related to the singularities of the
logarithm located at the points
\begin{eqnarray}\label{spoints}
\tau_{\pm} = \frac{r}{\bar{v}_+}\left(\Omega \pm i\sqrt{1 -
\Omega^2}\right)\,, \quad \Omega =
\frac{(\bm{r}\cdot\bar{\bm{v}}_+)}{r\bar{v}_+}
\end{eqnarray}
\noindent in the complex $\tau$-plane. For this purpose, we first
formally integrate Eq.~(\ref{kernel1}) by parts
\begin{eqnarray}
K(x,y,\tilde{x},t)&\stackrel{\circ}{=}& -
\frac{1}{\pi}\int\limits_{0}^{T+t}d
\left(\int\limits_{\tau}^{T+t}d\tau_1
M(\tilde{x},t-\tau_1)\right)\ln\left\{ \bar{v}_+^2 \tau^2 - 2
(\bm{r}\cdot \bar{\bm{v}}_+)\tau + r^2\right\}^{1/2} \nonumber
\\& = & \frac{\ln r}{\pi} \int\limits_{0}^{T+t}d\tau_1
M(\tilde{x},t-\tau_1)\nonumber\\
&&\qquad + \frac{1}{2\pi}\int\limits_{0}^{T+t}d\tau
\left(\int\limits_{\tau}^{T+t}d\tau_1
M(\tilde{x},t-\tau_1)\right)\left\{\frac{1}{\tau - \tau_+ } +
\frac{1}{\tau - \tau_- }\right\}\,. \nonumber
\end{eqnarray}
\noindent The first term on the right gives rise to a pure potential
which is bounded, so it can be omitted. Denoting also
\begin{eqnarray}\label{eumdef}
\int\limits_{\tau}^{T+t}d\tau_1 M(\tilde{x},t-\tau_1) \equiv
\EuScript{M}(\tilde{x},\tau,t)\,,
\end{eqnarray}
\noindent we thus have
\begin{eqnarray}&&
K(x,y,\tilde{x},t) \stackrel{\circ}{=}
\frac{1}{2\pi}\int\limits_{0}^{T+t}d\tau
\EuScript{M}(\tilde{x},\tau,t)\left\{\frac{1}{\tau - \tau_+ } +
\frac{1}{\tau - \tau_- }\right\}\,.\nonumber
\end{eqnarray}
\noindent To extract the singular part of this integral, we deform
the contour of integration in the complex $\tau$-plane so as to move
it away from the poles; this of course implies certain time-wise
restrictions on the function $\EuScript{M}(\tilde{x},\tau,t),$ and
hence on $M(\tilde{x},t).$ We shall return to this matter later (see
Sec.~7 and Appendix). Here we mention only that in essence, the
function $M(\tilde{x},t)$ is required to be analytic in a vicinity
of the real axis in the complex $t$-plane, which is guaranteed by
the existence of nonzero $\lambda_c.$ To respect reality of the
kernel, we take the singular part as a half sum of two expressions
obtained respectively by deforming the contour above and below the
real axis. The contribution we are interested in comes from moving
the contours beyond the poles (\ref{spoints}) [see Fig.~\ref{fig2}].

According to the Cauchy theorem,
\begin{eqnarray}\label{kernel2}
K(x,y,\tilde{x},t) & \stackrel{\circ}{=} & \frac{2\pi
i}{4\pi}\{\EuScript{M}(\tilde{x},\tau_+,t) - \EuScript{M}(\tilde{x},\tau_-,t)\}\nonumber\\
&& \qquad + \frac{1}{4\pi}\int\limits_{C_- \cup C_+}d\tau
\EuScript{M}(\tilde{x},\tau,t)\left\{\frac{1}{\tau - \tau_+ } +
\frac{1}{\tau - \tau_- }\right\}\,.
\end{eqnarray}
\noindent Instead of proving that the integral over the contour $C_-
\cup C_+$ is free of singularity and gives rise to a bounded
divergence-free potential, it is easier to show that the first term
in Eq.~(\ref{kernel2}) is bounded and reproduces correctly the
vorticity distribution at the flame front. This is done in Appendix.
Thus,
\begin{eqnarray}\label{kernel3}&&
K(x,y,\tilde{x},t) \stackrel{\circ}{=} \frac{i}{2}
\int\limits_{\tau_+}^{\tau_-}d\tau M(\tilde{x},t-\tau) \,.
\end{eqnarray}
\noindent Inserting Eq.~(\ref{kernel3}) into Eq.~(\ref{vint2v})
yields finally
\begin{eqnarray}\label{vortexp}
v^v_i \stackrel{\circ}{=} \frac{i}{4}\varepsilon_{ik}\partial_k
\int\limits_{-A}^{+A}d\tilde{x} \int\limits_{\tau_-}^{\tau_+}d\tau
M(\tilde{x},t-\tau)\,.
\end{eqnarray}
\noindent Since $\tau^*_+ = \tau_-,$ the right hand side of this
relation is real.

To make the meaning of the calculation performed more vivid
qualitatively, it is useful to mention an interrelation between the
roles the conditions a)--c) play in the above derivation. When
calculating the near-the-front value of the kernel
$K(x,y,\tilde{x},t),$ we retain terms of the first order with
respect to $\tau.$ This is sufficient for the calculation of the
vortex component of velocity at the front, taking into account that
this quantity is determined by the first spatial derivatives of the
kernel, and that $r=\tau v_+$ at the point defined by
Eq.~(\ref{point1}). This means, in particular, that $\tau$ in the
function $M(\tilde{x},t-\tau)$ cannot be neglected. As was mentioned
above, this function cannot be expanded in $\tau$ either without
violating c): the condition $M(\tilde{x},t)\to 0$ for $t\to -\infty$
guarantees convergence of the $\tau$-integral in the limit
$T\to\infty.$ At the same time, it is seen from Eq.~(\ref{vortexp})
that as a result of the $\tau$-integration, dependence of the
function $M(\tilde{x},t-\tau)$ on $\tau$ is transmuted into
coordinate dependence. This dependence does not affect the vorticity
distribution along the front, because $\tau = r/v_+$ along the
streamlines, so that $\tau = 0$ when the observation point is taken
on the front ($r=0$). Thus, the seemingly innocent condition c)
entails a nontrivial change in the structure of the potential
component of the burnt gas velocity in comparison with the steady
regime. In the latter case, on the other hand, the condition
$M(\tilde{x},t)\to 0$ for $t\to -\infty$ does not apply, but since
the $M$-function is independent of time, it remains independent of
the coordinates $(x,y)$ at all stages of the calculation. Proceeding
then as in Ref.~\citep{kazakov2}, one can verify that the divergent
contribution to the velocity field, coming from the integration over
large $\tau,$ is also coordinate-independent, so that the condition
c) is still met. Finally, it is not difficult to show that the
expression (\ref{vortexp}) cannot be further simplified following
the lines of Ref.~\citep{kazakov2} by omitting the additional
potential contribution {\it after} the spatial differentiation: it
turns out that this contribution satisfies the condition a) only in
the steady case. We shall return to this point later in
Sec.~\ref{ldinstab}.

\subsection{Definition of the vorticity mode}\label{vortdefinition}

Having obtained an explicit expression for the vortex component of
the burnt gas velocity for a finite $\Sigma,$ we have to consider
the question of the transition to the limits $A \to \infty,$ $T \to
\infty.$ Generally, the rule these limits are to be taken depends on
the problem under consideration. In the case of unsteadily
propagating flames, this issue is complicated by the fact that the
expression found for the vortex component is essentially nonlocal,
both in space and time. The latter nonlocality shows itself
explicitly through the $\tau$-integration in Eq.~(\ref{vortexp}),
and is to be expected from the very outset. In fact, appearance of
time non-locality is {\it inevitable}, because there must exist some
mechanism transferring the influence of the flame history onto its
current state. Such mechanism is unnecessary only in the stationary
case, where all information about the flame past is in a sense left
at the infinity downstream. Furthermore, we have seen that the time
dependence of the function $M(\tilde{x},t)$ is partially transmuted
into coordinate dependence, so the time non-locality entails
naturally an essential spatial non-locality. This is again in
contrast with the steady case, where it turned out to be possible to
find a local on-shell expression for the vortex component \citep[Cf.
Eq.~(37) in][]{kazakov2}.

Although the parameter $T$ does not appear explicitly in
Eq.~(\ref{vortexp}), it should be kept in mind that in the course of
derivation of this expression, the time dependence of the
$M(\tilde{x},t)$ function has transmuted into dependence on the
spatial coordinates. Hence, in order to preserve property c) of the
potential component, the limit $T\to \infty$ is generally to be
taken under assumption of vanishing of the function $M(\tilde{x},t)$
for $t\to -\infty.$ As to the limit $A\to \infty,$ any such
assumption would be irrelevant because of the flame periodicity
along the $x$-axis. To ensure convergence in this case, we introduce
an intermediate regularization of the $x$-integral replacing
Eq.~(\ref{vortexp}) by
\begin{eqnarray}\label{vortdefr}
v^v_i(\mu) \stackrel{\circ}{=} \frac{i}{4}\varepsilon_{ik}
\int\limits_{-A}^{+A}d\tilde{x}e^{-\mu r}\frac{\partial}{\partial
x_k} \int\limits_{\tau_-}^{\tau_+}d\tau M(\tilde{x},t-\tau)\,,
\end{eqnarray}
\noindent where $\mu >0$ is a sufficiently large parameter. We then
take the limit $A \to \infty$ and {\it define} the vorticity mode as
{\it the analytic} continuation of (\ref{vortdefr}) to the value
$\mu = 0$ along the real axis in the complex $\mu $-plane. Replacing
also the symbol $\stackrel{\circ}{=}$ by the equality sign, the
definition thus reads
\begin{eqnarray}\label{vortdef}
v^v_i = \frac{i}{4}\varepsilon_{ik} \left\{
\int\limits_{-\infty}^{+\infty}d\tilde{x}e^{-\mu
r}\frac{\partial}{\partial x_k} \int\limits_{\tau_-}^{\tau_+}d\tau
M(\tilde{x},t-\tau)\right\}_{\mu = 0^+}\,.
\end{eqnarray}
\noindent Let us show that the given definition using analytic
continuation in $\mu$ respects the properties a) -- c). Rewriting
Eq.~(\ref{vortdef}) as
\begin{eqnarray*}
 v^v_i & = & \frac{i}{4}\varepsilon_{ik}
\int\limits_{-A_0}^{+A_0}d\tilde{x}\frac{\partial}{\partial x_k}
\int\limits_{\tau_-}^{\tau_+}d\tau M(\tilde{x},t-\tau) \\
&& \qquad + \frac{i}{4}\varepsilon_{ik} \left\{
\left[\int\limits_{-\infty}^{-A_0} +
\int\limits_{+A_0}^{+\infty}\right] d\tilde{x}e^{-\mu
r}\frac{\partial}{\partial x_k} \int\limits_{\tau_-}^{\tau_+}d\tau
M(\tilde{x},t-\tau)\right\}_{\mu = 0^+}\,,
\end{eqnarray*}
where $A_0>0$ is arbitrary, and comparing with Eq.~(\ref{vortexp})
one sees that taking the limit $A \to \infty$ followed by the
analytic continuation in $\mu $ does not change the vorticity
distribution in the arbitrarily large domain $|x|< A_0.$ Therefore,
the above analytical operations amount to addition of some potential
field, so that b) is met. Furthermore, this field is bounded in the
sense of c). To see this, consider the quantity
\begin{eqnarray}\label{quant}
\frac{\partial}{\partial x_k} \int\limits_{\tau_-}^{\tau_+}d\tau
M(\tilde{x},t-\tau)\,.
\end{eqnarray}
\noindent Using Eq.~(\ref{spoints}) and the Newton-Leibnitz formula,
this expression is a combination of the functions
$M(\tilde{x},t-\tau_{\pm})$ times spatial derivatives of
$\tau_{\pm}.$ It follows from Eq.~(\ref{spoints}) that $|\tau_{\pm}|
= r/\bar{v}_+ \to \infty$ for $|\tilde{x}|\to \infty.$ Hence, if we
assume that $M(\tilde{x},t)$ is exponentially bounded in a vicinity
of the point $t=\infty$ in the complex $t$-plane, {\it i.e.,}
$|M(\tilde{x},t)| < e^{c |t|}$ for some $c>0$ and $|t|\to \infty,$
then there exists a large enough $\mu$ such that the
$\tilde{x}$-integral in Eq.~(\ref{vortdef}) converges. On the other
hand, the function $M(\tilde{x},t)$ is periodic with respect to
$\tilde{x}$ as the result of the flow periodicity. Therefore, all
singularities of the expression in the curly brackets in
Eq.~(\ref{vortdef}) are off the real axis in the complex
$\mu$-plane, except possibly for a simple pole at $\mu = 0.$ The
latter corresponds to an additive constant, $B_k,$ in the quantity
(\ref{quant}). The appearance of such a term is not forbidden by the
requirement of periodicity in $\tilde{x}.$ $B_k$ is independent of
$\bm{x}$ by virtue of the same flow periodicity. But after the
$\tilde{x}$-integration this term gives rise to a contribution of
the form $B_k/\mu + h_k(x),$ where $h_k(x)$ vanishes for $\mu\to 0.$
Since $B_k/\mu$ disappears upon spatial differentiation, $Dv^v_i$
can be continued analytically to $\mu = 0,$ so the property c) is
fulfilled indeed. Finally, the divergence of the vorticity mode
$${\rm div}\,\bm{v}^v = -
\frac{i}{4}\left\{
\mu\int\limits_{-\infty}^{+\infty}d\tilde{x}e^{-\mu
r}\frac{r_i}{r}\varepsilon_{ik}\frac{\partial}{\partial x_k}
\int\limits_{\tau_-}^{\tau_+}d\tau M(\tilde{x},t-\tau)\right\}_{\mu
= 0^+}$$ is proportional to $\mu.$ On the other hand, one has $r_i/r
= -{\rm sign}(\tilde{x})\delta_{1i} + O(1/|\tilde{x}|).$ Therefore,
the only term contributed by the integral, that survives after
continuation to $\mu = 0,$ is an $\bm{x}$-independent constant
proportional to $(\delta_{1i}\varepsilon_{ik}B_k/\mu).$ Thus, ${\rm
div}\,\bm{v}^v = {\rm const},$ and the condition a) is satisfied.

\section{Closed description of non-stationary flames}\label{closed}

We are now in position to write down an integro-differential
equation relating the on-shell values of the fuel velocity and the
front position function. Let us introduce the complex variable $z =
x + iy,$ and the complex velocity $\omega = u + i w.$ By virtue of
the properties a), b) the complex quantity $d\omega^p/dz,$ where
$\omega^p = u^p + i w^p,$ is an analytical function of the complex
variable $z$ in the downstream region. In conjunction with the
property c), analyticity of $d\omega^p/dz$ can be expressed in the
form of the following dispersion relation \citep{kazakov1,kazakov2}
\begin{eqnarray}\label{ch}
\left(1 + i \hat{\EuScript{H}}\right)\left(\omega^p_+\right)' = 0\,,
\end{eqnarray}
\noindent where the prime denotes $x$-differentiation, and the
action of the operator $\hat{\EuScript{H}}$ on an arbitrary function
$a(x)$ is defined by
\begin{eqnarray}\label{eq:4.2}
\left(\hat{\EuScript{H}}a\right)(x) = \frac{1 + i
f'(x,t)}{\pi}~\fint\limits_{-\infty}^{+\infty}
d\tilde{x}~\frac{a(\tilde{x})}{\tilde{x} - x + i[f(\tilde{x},t) -
f(x,t)]}\,,
\end{eqnarray}
\noindent slash denoting the principle value of the integral.
$\hat{\EuScript{H}}$ has properties similar to the Hilbert operator
$\hat{H}$ (and is effectively the Hilbert transform along the
front). In particular, it was proved by \cite{kazakov2} that
\begin{eqnarray}\label{pr}
\hat{\EuScript{H}}^2 = - 1\,.
\end{eqnarray}
\noindent The identity (\ref{ch}) relates in a complicated way the
on-shell values of the burnt gas velocity and the flame front
position. The fuel velocity also satisfies the conditions a) -- c),
this time in the upstream region. Indeed, a) is just the
differentiated Eq.~(\ref{flow1}), b) follows from the Thomson
theorem and the boundary conditions upstream, and c) is true because
the fuel velocity is bounded. The consequence of these properties is
the following dispersion relation for $\omega_- = u_- + iw_-$
\begin{eqnarray}\label{chup}
\left(1 - i \hat{\EuScript{H}}\right)\left(\omega_-\right)' = 0\,.
\end{eqnarray}
\noindent Denote $[\bm{v}]$ the jump of the gas velocity across the
flame front, $[\bm{v}] = \bm{v}(x,f(x,t)+0) - \bm{v}(x,f(x,t)-0).$
Then the sought equation for $\omega_-,f$ is obtained by
substituting
$$\omega^p_+ = - \omega^v_+ + \omega_- + [\omega]$$ in Eq.~(\ref{ch}),
and using Eqs.~(\ref{vortdef}), (\ref{chup})
\begin{eqnarray}\label{generalc1}&&
2\left(\omega_-\right)' + \left(1 +
i\hat{\EuScript{H}}\right)\left\{[\omega] - \frac{i}{4}
\int\limits_{-\infty}^{+\infty}d\tilde{x}e_k\frac{\partial}{\partial
x_k} \int\limits_{\tau_-}^{\tau_+}d\tau M(\tilde{x},t-\tau)\right\}'
= 0\,,
\end{eqnarray}
\noindent where $e_k = \varepsilon_{2k} + i\varepsilon_{1k},$ and we
omit for brevity the regularizing factor $e^{-\mu r}$ in the
integrand as well as the accompanying symbol of analytic
continuation. In the last term on the left, the argument $y$ is
understood to be set equal to $f(x,t)$ after the spatial partial
differentiation is performed, but before the $x$-differentiation
denoted by the prime. The value of vorticity at the front and the
normal velocity of the burnt gas, entering the function
$M(\tilde{x},t-\tau),$ as well as the velocity jumps at the front
are known functionals of the on-shell fuel velocity
\citep{matalon,pelce}. For instance, for zero-thickness flame fronts
one has
\begin{eqnarray}\label{jumps}
\bar{v}^n_+ &=& \theta\,, \quad [u] = \frac{\theta - 1}{N}\ ,
\quad [w] = - f'\frac{\theta - 1}{N}\,, \\
\sigma_+ &=& - \frac{\theta - 1}{\theta N}\left\{\frac{Dw_-}{Dt} +
f' \frac{Du_-}{Dt} + \frac{1}{N}\frac{Df'}{Dt}
\right\}\,,\label{vorticity}
\end{eqnarray}
\noindent where
$$\frac{D}{Dt} \equiv \frac{\partial}{\partial t} + \left(w_- +
\frac{f'}{N}\right)\frac{\partial}{\partial x}\,.$$ Thus, the
complex Eq.~(\ref{generalc1}) gives two equations for three
functions $w_-(x,t),$ $u_-(x,t)$ and $f(x,t).$ Together with the
evolution equation
\begin{eqnarray}\label{evolutiongen}
(\bar{\bm{v}}_-\cdot \bm{n}) = 1 + S(u_-,w_-,f')\,,
\end{eqnarray}
\noindent where $S$ is a known functional of its arguments,
proportional to the flame front thickness (or, rather, the cut-off
wavelength $\lambda_c$), Eq.~(\ref{generalc1}) provides one with a
closed description of unsteady flames in the most general form. Its
application to various particular problems is given in the next
section.

Before proceeding, the following point is worth to comment on. In
the analysis that led us to (\ref{generalc1}), the front was
represented by a graph $y=f(x,t),$ which excludes the overhangs or
the fronts that double back on themselves. This over-restrictive
assumption, adopted so far for simplicity, can be relaxed as
follows. Let us parameterize the flame front by a real parameter
$\xi$ so that $(x(\xi),y(\xi))$ be a diffeomorphic mapping of the
interval $-\infty < \xi < +\infty$ onto the front at time instant
$t.$ Then, setting $x(\xi)+i y(\xi) = Z(\xi,t),$ we define the new
metric coefficient $N(\xi,t)$ in terms of the infinitesimal
arclength along the front, $dl,$ by $dl = N(\xi,t)\, d\xi$; in other
words, $N(\xi,t) = \left|
\partial_{\xi} Z(\xi,t)\right|,$ which is nowhere singular for smooth fronts.
As long as the fuel and the burnt-gas regions remain connected
domains, the Cauchy theorem guarantees the existence of a
generalized operator $\hat{\EuScript{H}}$ such that $(1\pm
i\hat{\EuScript{H}})d\omega_{\pm}/d\xi = 0$. Specifically, when
acting on a smooth $a(\xi,t)$, the new $\hat{\EuScript{H}}$ produces
\begin{equation}
\left(\hat{\EuScript{H}}a\right)(\xi,t) =
\frac{\partial_{\xi}Z}{\pi}
\,\fint\frac{a(\tilde{\xi},t)~d\tilde{\xi}}{Z(\tilde{\xi},t)-Z(\xi,t)},
\end{equation}
instead of (\ref{eq:4.2}).  Accordingly, if $M(\xi,t)$ is still
defined as $M=N(\xi,t)\,\sigma_+(\xi,t)\,\bar{u}^n_+(\xi,t)$,
Eq.~(\ref{generalc1}) is formally unchanged, except that the prime
now denotes $d/d\xi.$ Of course, the time derivatives, now at fixed
$\xi$, must be handled in a way consistent with the new
representation of the flame front. Yet such a re-parameterization
does not capture situations when isolated pockets of unburnt fuel
form, because the fresh domain then ceases to be path-connected.
Unfortunately, as long as a {\it local} propagation law [Cf.
(\ref{evolutiongen})] is employed such a phenomenon cannot be
excluded {\it a priori}: in no way can a flame element ``know'' that
another one is to produce a ``head-on'' collision.

The above re-parameterization is unnecessary for the wrinkled fronts
considered below.

\section{Equations~(\ref{generalc1}), (\ref{evolutiongen}) in limiting cases
}\label{verification}

To give a consistency check for Eq.~(\ref{generalc1}) and also to
gain a deeper insight into the structure of this equation, we use it
below to derive anew classical results on flame front dynamics.

\subsection{Darrieus-Landau instability of zero-thickness
flames}\label{ldinstab}

Let us consider first the classical linear stability problem of
zero-thickness planar flame propagation \citep{darrieus,landau}. In
this case, $\bar{v}^n_+ = \theta,$ $N=1,$ while the linearized
on-shell vorticity (\ref{vorticity}) reads
$$\sigma_+ = - \frac{\theta - 1}{\theta}
\left(\frac{\partial w_-}{\partial t} + \frac{\partial^2 f
}{\partial t \partial x}\right)\,.$$ Accordingly, expression
(\ref{vortdef}) simplifies to
\begin{eqnarray}\label{vortdefl}
v^v_i = - \frac{i(\theta - 1)}{4}\varepsilon_{ik} \left\{
\int\limits_{-\infty}^{+\infty}d\tilde{x}e^{-\mu
|x-\tilde{x}|}\frac{\partial}{\partial x_k}
\int\limits_{\tau_-}^{\tau_+}d\tau \left(\frac{\partial
w_-}{\partial t} + \frac{\partial^2 f }{\partial t \partial
x}\right)\right\}_{\mu = 0^+}\,.
\end{eqnarray}
\noindent Since the integrand here is a first order quantity, it is
sufficient to calculate $\tau_{\pm}$ for a plane front which is
assumed to be at $y=0$
\begin{eqnarray}\label{spointsld}
\tau_{\pm} = \frac{y}{\theta} \pm i \frac{|x-\tilde{x}|}{\theta}\,.
\end{eqnarray}
\noindent Let the disturbance be periodic in $x,$ growing
exponentially with time. Spatial periodicity of the linear problem
is most conveniently represented in the complex form, in which case
\begin{eqnarray}\label{eq:5.3}
f(x,t), u_-(x,t), w_-(x,t) \sim e^{ikx + \nu t}\,.
\end{eqnarray}
\noindent However, it should be kept in mind that the coefficients
in Eq.~(\ref{generalc1}) are also complex. In order to preserve the
right complex structure of this equation, all the functions involved
are to be written in the form containing no imaginary coefficients.
To find $v^v_i,$ we have to evaluate the following integrals
$$I_k(x,y,t,\mu) = \int\limits_{-\infty}^{+\infty}d\tilde{x}e^{-\mu |x-\tilde{x}|}\frac{\partial}{\partial x_k}
\int\limits_{y/\theta-i|x-\tilde{x}|/\theta}^{y/\theta+i|x-\tilde{x}|/\theta}d\tau
e^{ik\tilde{x} + \nu (t - \tau)}$$ for $\mu >0$ and $k=1,2.$ A
straightforward calculation gives
\begin{eqnarray*}
I_1(x,y,t,\mu ) & = & \frac{i}{\theta}e^{ikx + \nu t - \nu y/\theta}
\left\{\left(\frac{1}{\mu + ik + i\nu/\theta} + \frac{1}{-\mu + ik +
i\nu/\theta}\right)
+ (\nu \rightarrow -\nu)\right\}\,,\\
I_2(x,y,t,\mu ) & = & \frac{1}{\theta}e^{ikx + \nu t - \nu y/\theta}
\left\{\left(\frac{1}{\mu + ik + i\nu/\theta} + \frac{1}{-\mu + ik +
i\nu/\theta}\right) - (\nu \rightarrow -\nu)\right\}\,.
\end{eqnarray*}
where ($\nu \rightarrow -\nu$) is shorthand for the preceding
parenthesis with $\nu$ changed into its opposite. The on-shell
values of these functions analytically continued to $\mu = 0$ are
\begin{eqnarray}\label{i1i2}
I_1(x,0,t,0) = \frac{4}{\theta}e^{ikx + \nu t} \frac{k}{k^2 -
\nu^2/\theta^2}\,, \quad I_2(x,0,t,0) = \frac{4i}{\theta}e^{ikx +
\nu t} \frac{\nu/\theta}{k^2 - \nu^2/\theta^2}\,.
\end{eqnarray}
\noindent Using this in Eq.~(\ref{vortdefl}) yields
\begin{eqnarray}\label{wvld}
w^v_+ &=& \frac{\nu/\theta}{k^2 - \nu^2/\theta^2} \frac{(\theta -
1)}{\theta}(\nu w_- + ik\nu f)\,,
\\
u^v_+ &=& \frac{ik}{k^2 - \nu^2/\theta^2} \frac{(\theta -
1)}{\theta}(\nu w_- + ik\nu f) \,.\label{uvld}
\end{eqnarray}
\noindent To put these expressions in an explicitly real form, it is
sufficient to write
\begin{eqnarray}
w^v_+ = - \frac{\sigma_+\nu/\theta}{k^2 - \nu^2/\theta^2}\,, \quad
u^v_+ = - \frac{\sigma'_+ }{k^2 - \nu^2/\theta^2} \,.\nonumber
\end{eqnarray}
\noindent Next, we note that in the linear approximation, the
operator $\hat{\EuScript{H}}$ becomes just the Hilbert operator
$\hat{H},$ whose action on the harmonic functions is defined by
\begin{eqnarray}\label{hilbert}
\hat{H}\exp(i k x) = i\chi(k)\exp(i k x)\,, \quad k\ne 0\,,
\end{eqnarray}
where
$$\chi(k) = \left\{
\begin{array}{cc}
+1,& k>0\,,\\
\phantom{+}0, & k = 0\,,\\
-1,&  k<0\,.
\end{array}
\right.
$$
Equation (\ref{generalc1}) thus becomes
\begin{eqnarray}\label{generalc1ld}&&
2(\omega_-)' + \left(1 + i\hat{H}\right)\left\{[\omega] +
\frac{\sigma'_+ + i \sigma_+ \nu/\theta}{k^2 -
\nu^2/\theta^2}\right\}' = 0\,.
\end{eqnarray}
\noindent Extracting the real and imaginary parts of this equation,
we find
\begin{eqnarray}\label{realimg}
2(u_-)' &+& \left\{[u] - \hat{H}[w] + \frac{\sigma'_+ -
\hat{H}\sigma_+
\nu/\theta}{k^2 - \nu^2/\theta^2}\right\}' = 0\,, \\
(w_-)' &=& \hat{H}(u_-)'\,.
\end{eqnarray}
\noindent Finally, upon linearization the jump conditions
(\ref{jumps}) simplify to
\begin{eqnarray}\label{jumpsld}
[u] = \theta - 1\,, [w] = - f'(\theta - 1)\,,
\end{eqnarray}
\noindent while the linearized evolution equation reads
\begin{eqnarray}\label{evolution}
u_- - \frac{\partial f}{\partial t} = 1\,.
\end{eqnarray}
\noindent Inserting these into Eq.~(\ref{realimg}) and then
substituting $\sigma_+ = - (\theta - 1)/\theta (i\chi(k)\nu^2 f +
ik\nu f)$ leads after some simple algebra to the equation
\begin{equation}\label{eq:5.12}
 \frac{\theta + 1}{\theta}\nu^2 + 2\nu |k| - (\theta - 1)k^2 = 0,
\end{equation}
which is nothing but the famous Darrieus-Landau dispersion relation
determining the perturbation growth rate as a function of the wave
number and the gas expansion coefficient \citep{darrieus,landau}
\begin{eqnarray}\label{lddisp}
\nu = \frac{\theta}{\theta + 1} \left(\sqrt{1 + \theta -
\frac{1}{\theta}} - 1\right)|k| .
\end{eqnarray}
The effects related to the finite front thickness can be explicitly
accounted for in the above linear analysis by including terms linear
in $f''$ in the right hand side of Eqs.~(\ref{jumpsld}),
(\ref{evolution}). This would changed the last term in
Eq.~(\ref{eq:5.12}) by the extra factor $(1- \left|k\right|
\lambda_c/2\pi),$ and make $\nu$ roughly parabolic for all
$\left|k\right|\le 2\pi/\lambda_c$; max($\nu$) enables one to
estimate the typical growth time as $t_{DL}\simeq 2\lambda_c/\pi
(\sqrt{\theta}-1)$, to be used later (Sec. 6).

Next, let us return to the remark made after Eq.~(\ref{kernel1}). It
was mentioned there that even if one is interested only in the
on-shell values of the vorticity component, the $\tau$-dependence of
the function $M(\tilde{x},t-\tau)$ cannot be neglected. Evidently,
doing so amounts simply to rewriting the equation obtained by
Kazakov \citep{kazakov1,kazakov2} for steady flames in terms of the
local current flow velocity relative to the front. It is not
difficult to verify that in the case under consideration, this would
change the coefficient of $\nu^2$ in Eq.~(\ref{eq:5.12}) to the
wrong value $(\theta-1)/\theta\,.$ This change is thus a reflection
of the memory effects encoded in the function $M(\tilde{x},t-\tau).$
Equation (\ref{generalc1}) properly takes into account these
effects, correctly reproducing the Darrieus-Landau relation, and
automatically captures all the aspects of flame dynamics related to
inertia.

At this stage of the analysis it is appropriate to pause and discuss
the meaning of the analytic continuation appearing in the definition
(\ref{vortdef}) in somewhat more detail. This continuation is used
to make the $\tilde{x}$-integral meaningful in the limit $A\to
\infty.$ One can avoid using this analytic means if the formal
result of improper integration along the infinite flame front is
treated in the sense of distributions. Indeed, in this case
expressions (\ref{i1i2}) for the vortex component would contain
additional terms proportional to $\delta(k + \nu/\theta)$ or
$\delta(k - \nu/\theta)$ coming from the integration over large
$\tilde{x}s.$ To see that these terms are inconsequential we recall
that the above consideration of the single $k$-mode evolution is not
completely adequate from the physical point of view, because in
practice one always deals with wave packets consisting of a
continuum of wave numbers $k,$ rather than with a single mode. This
means that the physical expression for the flame front position with
the given $k_0$ is obtained by integrating the found solution
$f(x,t)$ with some weight over a small but finite range $\Delta k$
around $k_0.$ Upon this integration all terms involving $\delta(k
\pm \nu/\theta)$ disappear, because $\nu = \pm k\theta$ are not
roots of the Darrieus-Landau relation. The two procedures are thus
equivalent.

\subsection{The small $(\theta - 1)$ expansion}

Let us next consider the case of small gas expansion. We will verify
that within the framework of the asymptotic expansion with respect
to $\theta - 1 \equiv \alpha,$ Eq.~(\ref{generalc1}) reduces at the
first post-Sivashinsky order to the well-known Sivashinsky-Clavin
equation \citep{sivclav}\footnote{It is worth noting that the small
expansion parameter used by Sivashinsky and Clavin is
$\gamma\equiv(\theta-1)/\theta = \alpha/(1+\alpha),$ rather than
$\alpha.$ The reason for switching from $\alpha$ to $\gamma,$ found
sometimes in the literature, is that the latter would improve the
expansion accuracy/convergence. The argument given in this
connection, namely, that $\gamma<1$ for all $\theta,$ while $\alpha$
is small only for $\theta$ close to $1,$ is not quite correct.
Validity of an expansion is determined not by the value of the
expansion parameter, but by the relative value of the terms
neglected in the course of the calculation. It is true that
expanding the decreasing function $f(\theta) = 1/\theta =
1/(1+\alpha)$ in terms of $\alpha/(1+\alpha)$ instead of $\alpha$ is
an improvement. However, it is not in the case of the increasing
function $f(\theta) = \theta - 1 = \alpha,$ which certainly appears
in the governing equations [see Eqs.~(\ref{generalc1}),
(\ref{jumps})]. Thus, whether a change of the expansion parameter
improves the expansion accuracy/convergence is a question of the
structure of the whole perturbation series, which cannot be resolved
from the knowledge of its first few terms. It is one of the goals of
our approach to make questions of this kind accessible for
theoretical analysis.}. To perform the asymptotic expansion we
recall that the cutoff wavelength for the short wavelength
perturbations $\lambda_c$ is of the order $1/\alpha.$ This means
that the wave numbers involved are $O(\alpha).$ In other words,
spatial differentiation of a flow variable raises its order by one;
in particular, $f' = O(\alpha).$ Also, Eq.~(\ref{lddisp}) tells us
that for small $\alpha,$ the perturbation growth rate $\nu =
k\alpha/2 = O(\alpha^2),$ so the order of a flow variable is raised
by two upon time differentiation. It follows then from
Eq.~(\ref{evolutiongen}) [with $S\equiv 0$] that $u_- = 1 +
O(\alpha^2),$ while potentiality of the upstream flow implies that
$w_- = O(\alpha^2)$ [this is clearly seen from the dispersion
relation (\ref{chup})]. The first post-Sivashinsky approximation
corresponds to retaining terms of the fourth order in
Eq.~(\ref{generalc1}), or equivalently, $O(\alpha^3)$-terms before
the spatial differentiation. It was shown by \cite{kazakov2} that to
this order, $\hat{\EuScript{H}}$ becomes just the Hilbert operator.
Therefore, like in the linear case considered above, the real and
imaginary parts of Eq.~(\ref{generalc1}) are readily separated to
give
\begin{eqnarray}\label{generalc1scu}
2\left(u_-\right)' &+& \left\{[u] - \hat{H}[w] +
(\varepsilon_{1k}\hat{H} - \varepsilon_{2k})\frac{i}{4}
\int\limits_{-\infty}^{+\infty}d\tilde{x}\frac{\partial}{\partial
x_k} \int\limits_{\tau_-}^{\tau_+}d\tau
M(\tilde{x},t-\tau)\right\}' = 0\,, \\
(w_-)' &=& \hat{H}(u_-)'\,.\label{generalc1scw}
\end{eqnarray}
\noindent Since the quantity $J_k\equiv
\partial_k\int_{\tau_-}^{\tau_+}d\tau M$ is under the sign of
spatial integral, we need the function $M(\tilde{x},t)$ to the
fourth order in $\alpha.$ To this accuracy, it is equal to the
on-shell vorticity
$$M = \sigma_+ = - (\theta - 1)\left(\dot{f}'
+ f'f''\right)\,,$$ the dot denoting differentiation with respect to
the time $t.$ As this quantity is already $O(\alpha^4),$
$\tau_{\pm}$ can be taken in the form (\ref{spointsld}) with $\theta
= 1.$ Thus, performing the spatial differentiation and setting $y =
0$ gives
\begin{eqnarray}
J_1 &=& i \chi(x-\tilde{x})
\left\{\sigma_+(\tilde{x},t-i|x-\tilde{x}|) +
\sigma_+(\tilde{x},t+i|x-\tilde{x}|)\right\}\,, \nonumber \\
J_2 &=& \left\{\sigma_+(\tilde{x},t-i|x-\tilde{x}|) -
\sigma_+(\tilde{x},t+i|x-\tilde{x}|)\right\}\,. \nonumber
\end{eqnarray}
\noindent Let us show that imaginary parts in the argument of
$\sigma_+$ can be omitted. Note that within the asymptotic expansion
in $\alpha,$ dependence of $\sigma_+$ on $i|x-\tilde{x}|$ can be
treated perturbatively. Indeed, let us write formally
$$\sigma_+(\tilde{x},t-i|x-\tilde{x}|) =
\sigma_+(\tilde{x},t) - i|x-\tilde{x}|\dot{\sigma}_+(\tilde{x},t) -
\frac{1}{2}(x-\tilde{x})^2\ddot{\sigma}_+(\tilde{x},t) + \cdots\,.$$
To assess the relative order of consecutive terms in this series, we
have first to get rid of the factors containing explicit coordinate
dependence, which can be done by successive integration by parts
with respect to $\tilde{x}.$ One has, for example,
\begin{eqnarray*}
\int\limits_{-\infty}^{+\infty} d\tilde{x}
|x-\tilde{x}|\dot{\sigma}_+(\tilde{x},t) & = &
\int\limits_{-\infty}^{+\infty} d \left(\int^{\tilde{x}}dx_1
\dot{\sigma}_+(x_1,t)\right)|x-\tilde{x}|,\\
& = & \int\limits_{-\infty}^{+\infty} d\tilde{x}\chi(x-\tilde{x})
\left(\int^{\tilde{x}}dx_1 \dot{\sigma}_+(x_1,t)\right)\,,
\end{eqnarray*}
 where
according to the discussion in the preceding section, the
contributions of infinitely remote parts of the front have been
omitted. Since each time differentiation adds two powers of
$\alpha,$ while spatial integration subtracts only one, we conclude
that the above expansion for $\sigma_+$ is effectively an asymptotic
series in powers of $\alpha.$ Hence, to the fourth order in $\alpha$
$$J_1 = 2i\chi(x-\tilde{x})
\sigma_+(\tilde{x},t)\,, \quad J_2 = 0\,,$$ and so
\begin{eqnarray*}
\int\limits_{-\infty}^{+\infty}d\tilde{x}J_1 & = & - 2i (\theta - 1)
\int\limits_{-\infty}^{+\infty}d\tilde{x}\chi(x-\tilde{x})\left(\dot{f}'
+ f'f''\right)(\tilde{x},t) \\ & = & - 4i(\theta - 1)
\int\limits_{-\infty}^{+\infty}d\tilde{x}\delta(x-\tilde{x})\left(\dot{f}
+ \frac{f'^2}{2}\right)(\tilde{x},t) \\ & = & - 4i(\theta - 1)
\left(\dot{f} + \frac{f'^2}{2}\right)(x,t)\,,
\end{eqnarray*}
\noindent where the boundary terms at $\tilde{x} = \pm \infty$ were
omitted again in the course of the integration by parts, and the
relation $\partial \chi(x-\tilde{x})/\partial\tilde{x} = -
2\delta(x-\tilde{x})$ was used. Inserting this together with the
velocity jumps into Eqs.~(\ref{generalc1scu}), (\ref{generalc1scw})
yields within the given accuracy
\begin{eqnarray}
2u_- + (\theta - 1)\left(\hat{H}f' + \dot{f} \right) =
C_1(t)\,,\\
2w_- + (\theta - 1)\left(-f' + \hat{H}\dot{f} \right) =
C_2(t)\,,\label{gen1w}
\end{eqnarray}
\noindent where $C_{1,2}(t)$ are two ``integration constants'' (some
$x$-independent functions of time). Since the left hand side of
(\ref{gen1w}) is odd in $x,$ one has $C_2 \equiv 0.$ Using these
formulas in the evolution equation (\ref{evolutiongen}) written in
the form
$$u_- - \dot{f} - f'w_- = 1 + \frac{f'^2}{2}$$
leads to the equation for the function $f(x,t)$
\begin{eqnarray}\label{sivclaveq}
\frac{\theta + 1}{2}\dot{f} + \frac{\theta}{2}f'^2 + C(t) = -
\frac{\theta - 1}{2}\hat{H}f'\,,
\end{eqnarray}
\noindent where $C(t) = 1 - C_1(t)/2.$ The function $C(t)$ can be
found by averaging the obtained equation along the flame front:
$$C(t) = - \frac{\theta + 1}{2}\langle\dot{f}\rangle - \frac{\theta}{2}
\langle f'^2 \rangle\,.$$ Equation (\ref{sivclaveq}) is nothing but
the Sivashinsky-Clavin equation \citep{sivclav}, with the term
$C(t)$ added according to \cite{joulin1}. Account of the transport
processes inside the front would have added a term proportional to
$\lambda_c f''$ to the right hand side of Eq.~(\ref{sivclaveq}).

Finally, it is interesting to note that the memory effects are
insignificant not only at the lowest order of the small
$(\theta-1)$-expansion, but also at the first post-Sivashinsky
approximation considered here. Namely, a direct calculation shows
that replacing $M(\tilde{x},t-\tau)$ by $M(\tilde{x},t)$ in
Eq.~(\ref{generalc1}) does not change Eq.~(\ref{sivclaveq}). This is
natural, because the condition $\alpha\ll 1$ implies slow dynamics.
However, this is not the case already in the next order, which is
possibly one reason why the memory effects are often overlooked.

\subsection{Flame propagation in time-dependent gravitational field}

Depending on the direction of flame propagation, gravitational field
leads either to strengthening or damping of the Darrieus-Landau
instability. Also, the influence of sound waves on the front
dynamics can be described effectively as the flame propagation in a
time-dependent gravitational field \citep{acoust}. Let $g(t)$ denote
its strength, with the convention that $g>0$ corresponds to a
stabilizing effect [see Eq.~(\ref{feqg})]. Inclusion of the
gravitational field does not change the flow equations
(\ref{flow1}), (\ref{flow2}), so that their consequence,
Eq.~(\ref{generalc1}), has the same structure. The jump conditions
(\ref{jumps}) for the gas velocity are also left intact by gravity.
The only place where $g(t)$ appears in our approach is the
expression of the on-shell vorticity (as a result of baroclinic
effects inside the front). Namely, the gravity-induced jump in this
quantity to be added to the right hand side of Eq.~(\ref{vorticity})
is \citep{hayes}\footnote{More generally, $\Delta\sigma_+ =
\frac{(\theta - 1)}{\theta N} \left\{\Gamma(x,f(x,t),t)\right\}'\,,$
where $\Gamma(x,y,t)$ is the gravitational potential, $\bm{g} = -
\nabla\Gamma.$}
\begin{eqnarray}\label{gvort}
\Delta\sigma_+ = -\frac{(\theta - 1)}{\theta N}g(t)f'(x,t)\,.
\end{eqnarray}
\noindent If development of the Darrieus-Landau instability is
suppressed by the gravitational field, then it is natural to
consider harmonic front perturbations instead of the exponentially
growing ones used in Sec.~\ref{ldinstab}. The choice between the two
representations is just a matter of convenience. Namely, it will be
seen below that although the intermediate procedure of analytic
continuation in $\mu$ is somewhat different for $\nu$ imaginary, the
final equations for the front position are simply analytic
continuations of one another with respect to frequency.

Because of the time dependence of $g(t),$ the function $f(x,t)$ is
now to be taken as a superposition of an arbitrary number of
harmonics
\begin{eqnarray}\label{fdecompose}
f(x,t) = \int\limits_{-\infty}^{+\infty}d\omega dk f(k,\omega)
e^{ikx - i\omega t}\,.
\end{eqnarray}
\noindent It is also convenient to introduce the Fourier
decomposition of the function $G(x,t) = g(t)f'(x,t)$
\begin{eqnarray}\label{gdecompose}
G(x,t) = \int\limits_{-\infty}^{+\infty}d\omega dk G(k,\omega)
e^{ikx - i\omega t}\,.
\end{eqnarray}
\noindent To perform the analytic continuation with respect to $\mu
$ in Eq.~(\ref{vortdef}), we will assume that $f(k,\omega),$
$G(k,\omega),$ considered as functions of $\omega,$ vanish outside a
large but finite frequency band $|\omega| \leqslant \omega_0.$ Then,
choosing $\mu > \omega_0/\theta,$ inserting the above Fourier
decompositions in the linearized expression for the on-shell
vorticity, performing $\tilde{x}$-integration in Eq.~(\ref{vortdef})
as before, and continuing the result analytically to $\mu = 0,$ we
find
\begin{eqnarray}
w^v_+ &=& -i\,\frac{\theta - 1}{\theta}
\int\limits_{-\infty}^{+\infty}d\omega dk\,e^{ikx - i\omega
t}\frac{\omega/\theta}{k^2 +
\omega^2/\theta^2}\left\{(-i\chi(k)\omega^2 + k\omega)f(k,\omega) +
G(k,\omega)\right\}\,, \nonumber
\\ \quad u^v_+ &=& i\,\frac{\theta - 1}{\theta}
\int\limits_{-\infty}^{+\infty}d\omega dk\,e^{ikx - i\omega t
}\frac{k}{k^2 + \omega^2/\theta^2}\left\{(-i\chi(k)\omega^2 +
k\omega)f(k,\omega) + G(k,\omega)\right\}\,.\nonumber
\end{eqnarray}
\noindent The first terms on the right of these equations are just
analytic continuations of the expressions (\ref{wvld}), (\ref{uvld})
to the imaginary value of the growth rate: $\nu \to -i\omega.$ As in
the Darrieus-Landau problem, Eq.~(\ref{generalc1}) splits into two
real equations
$$ 2(u_-)' + \left\{[u] - \hat{H}[w] - u^v_+ + \hat{H}w^v_+ \right\}' = 0\,, \\
\quad (w_-)' = \hat{H}(u_-)'\,,$$ which after substitution of the
found expressions for $v^k_i$ together with Eqs.~(\ref{jumpsld}),
(\ref{evolution}) yield
\begin{eqnarray}&&
\int\limits_{-\infty}^{+\infty}d\omega dk\,e^{ikx - i\omega
t}\left\{\left[\frac{\theta + 1}{\theta}\omega^2 + 2i\omega |k| +
(\theta - 1)k^2\right]k f(k,\omega) \right.\nonumber \\&&
\hspace{4cm}\left. + \frac{\theta - 1
}{\theta}ik\chi(k)G(k,\omega)\right\}\left(\frac{\omega}{\theta} +
ik\right)^{-1} = 0\,.\nonumber
\end{eqnarray}
\noindent Acting on this equation by the operator
$$\left(\frac{i}{\theta}\frac{\partial}{\partial t} +
\frac{\partial}{\partial x}\right)\int dx,$$ and taking into account
definitions (\ref{fdecompose}), (\ref{gdecompose}), we arrive at the
well-known equation for the front position \citep{acoust}
\begin{eqnarray}\label{feqg}
\frac{\theta + 1}{\theta}\ddot{f} - 2\hat{H}\dot{f}' + (\theta -
1)f'' - \frac{\theta - 1 }{\theta}g(t)\hat{H}f' = C(t)\,,
\end{eqnarray}
\noindent where $C(t)$ is a function of time appearing as the result
of the spatial integration symbolized by $\int dx.$ As before,
$C(t)$ can be found by averaging Eq.~(\ref{feqg}) along the front.
Equation (\ref{feqg}) is well-known to be the key ingredient in
understanding the parametric flame response to oscillating $g(t)$s
\citep{markstein1951,searby1991}, in which case the inertia (hence,
memory) effects play the central role.

\subsection{Steady flame propagation}

In the case of stationary flames, the $M$-function is
time-independent, whence the $\tau$-integration in
Eq.~(\ref{generalc1}) is trivial. One has
\begin{eqnarray*}
\frac{i}{4}
\int\limits_{-\infty}^{+\infty}d\tilde{x}e_k\frac{\partial}{\partial
x_k} \int\limits_{\tau_-}^{\tau_+}d\tau M(\tilde{x},t-\tau) & = & -
\frac{1}{2}
\int\limits_{-\infty}^{+\infty}d\tilde{x}\frac{M(\tilde{x})}{v_+}
e_k\frac{\partial}{\partial x_k}\sqrt{r^2 -
\frac{(\bm{r}\cdot\bm{v}_+)^2}{v^2_+}},\\
& = & - \frac{1}{2}
\int\limits_{-\infty}^{+\infty}d\tilde{x}M(\tilde{x})
\frac{e_k\beta_k}{v_+}\,,
\end{eqnarray*}
 where
\begin{eqnarray}\label{vectorbeta}
\beta_k = \left(\frac{r_k}{r} - \Omega\frac{v_{k+}}{v_+}\right)
\frac{1}{\sqrt{1-\Omega^2}}\,.
\end{eqnarray}
\noindent This vector satisfies
$$\beta_k\beta_k = 1, \qquad \beta_k v_{k+} = 0\,,$$
i.e., $\bm{\beta}$ is the unit vector orthogonal to ${\bf v}_+\,.$
In addition to that, $\bm{\beta}$ changes its sign at the point
$\tilde{x}$ satisfying Eq.~(\ref{point1}). Hence, its scalar product
with the complex vector $e_i$ can be written on-shell as
$$e_k\beta_k = - \frac{\omega_+}{v_+}\chi(x - \tilde{x})\,.$$
Differentiation with respect to $x$ using the formula $\chi'(x) =
2\delta(x)$ makes the $\tilde{x}$-integration trivial, so that
Eq.~(\ref{generalc1}) takes the form
\begin{eqnarray}\label{generalc1st}&&
2\left(\omega_-\right)' + \left(1 +
i\hat{\EuScript{H}}\right)\left\{[\omega]' -
\frac{Nv^n_+\sigma_+\omega_+}{v^2_+} \right\} = 0
\end{eqnarray}
\noindent which is exactly Eq.~(42) of Ref.~\cite{kazakov2}.

In connection with Eq.~(\ref{generalc1st}) one can indicate why
viscous effects have virtually no influence on the shape and the
speed of steady flames, except possibly a small one on the cut-off
wavelength ($\lambda_c$) itself, the only internal reference length
of our problem. The Reynolds number based on $\lambda_c$ and the
{\it burnt} gas properties and speeds is normally about $10^2.$ The
viscous length $\lambda_{\mathrm{vis}}$ -- the distance at which the
vorticity present behind the front-crests of transverse size $\sim
\lambda_c$ dissipates noticeably -- thus exceeds $\lambda_c$ by two
orders. This is definitely too late (in terms of the Lagrangian time
$\tau$) to modify Eq.~(\ref{generalc1st}), whose validity only
requires that $\sigma\rightarrow \sigma_+$ in an infinitesimal layer
($\tau \rightarrow 0^+$) downstream of the front ($\tau=0$). As to
unsteady flames the picture is more involved, even though the
potential fuel flow is still unaffected directly by viscous effects.
The problem would certainly deserve further studies, for one cannot
{\it a priori} exclude viscosity-related large-scale phenomena
occurring far from the front, that could nevertheless influence its
dynamics indirectly. We merely mention here that the spontaneous
flame dynamics is initially little affected, because the
Darrieus-Landau time $t_{DL}$ (see Sec. 5.1) is well shorter than
$t_{\mathrm{vis}} = \lambda_{\mathrm{vis}}/\theta$: the decay of
$M(\tilde{x},t-\tau)$ for $\tau \gg 1$ is fully controlled by the
wrinkle growth itself.

\section{Stability analysis of general steady flame patterns}

Let us use Eq.~(\ref{generalc1}) to derive an equation governing
propagation of small disturbances along a given steady flame
pattern. As in the Darrieus-Landau or stationary problems considered
above, Eq.~(\ref{generalc1}) essentially simplifies in this case,
because the time non-locality is no longer a complication. Indeed,
since there is no external time-varying field (as the stationary
regime is assumed to exist) and the flame disturbance is small, it
is sufficient to consider perturbations having the form
$$\delta f(x,t) = \tilde{f}(x)e^{\nu t}\,, \quad
\delta w_-(x,t) = \tilde{w}(x)e^{\nu t}\,, \quad \delta u_-(x,t) =
\tilde{u}(x)e^{\nu t}\,,$$ in which case the time dependence is
prescribed, and the $\tau$-integration in Eq.~(\ref{generalc1}) is
readily done. Let us gather the functions $f(x,t),$ $w_-(x,t),$
$u_-(x,t)$ into an array $\{\xi_{\alpha}(x,t)\},$ $\alpha = 1,2,3$:
$$(\xi_1,\xi_2,\xi_3) = (f,w_-,u_-)\,,$$ and denote
$\xi^{(0)}_{\alpha}(x)$ the given solution of
Eq.~(\ref{generalc1st}). We have to linearize the left hand side of
Eq.~(\ref{generalc1}) with respect to $\delta\xi(x,t) \equiv
\tilde{\xi}(x)e^{\nu t} = (\tilde{f},\tilde{w},\tilde{u})e^{\nu t}.$
Depending on whether the function $M(\tilde{x},t)$ is varied, one
finds contributions of two types. If $\tilde{\xi}_{\alpha}$ comes
from terms other than $M,$ then $M(\tilde{x},t)$ is evaluated for
$\xi = \xi^{(0)}$ and hence is time-independent. As we saw in the
preceding section, Eq.~(\ref{generalc1}) simplifies in this case to
Eq.~(\ref{generalc1st}), so the corresponding contribution to the
variation is given by the left hand side of Eq.~(\ref{generalc1st})
linearized around $\xi^{(0)},$ with $M$ kept fixed. The other
contribution comes from variation of $M(\tilde{x},t)$ and has the
form
$$\Delta_M = \left(1 +
i\hat{\EuScript{H}}\right)\left\{-\frac{i}{4}
\int\limits_{-\infty}^{+\infty}d\tilde{x}e_k\frac{\partial}{\partial
x_k} \int\limits_{\tau_-}^{\tau_+}d\tau
\hat{M}_{\alpha}({\tilde{x}})\tilde{\xi}_{\alpha}(\tilde{x})
e^{\nu(t-\tau)}\right\}'\,,$$ where $\hat{M}_{\alpha}({\tilde{x}})$
is the differential operator obtained by linearizing the function
$M(\tilde{x},t)$ around the stationary solution, and changing
$\partial/\partial t \to \nu$ afterwards. A straightforward
calculation gives
$$\Delta_M = - \frac{e^{\nu t}}{2}\left(1 +
i\hat{\EuScript{H}}\right)
\int\limits_{-\infty}^{+\infty}d\tilde{x}\hat{M}_{\alpha}({\tilde{x}})
\tilde{\xi}_{\alpha}(\tilde{x})\frac{\omega_+}{v^2_+}
\left\{\exp\left(- \frac{\nu r}{v_+}e^{-i\phi}\right)\chi(x -
\tilde{x})\right\}'\,,$$ where $\phi\in [-\pi,+\pi]$ is the angle
between the vectors $\bm{r},$ $\bm{v}_+,$ defined positive if the
rotation from $\bm{v}_+$ to $\bm{r}$ is clockwise. It is not
difficult to verify that the $x$-differentiation of the step
function in the latter expression gives rise to the term which is
just the variation of the left hand side of Eq.~(\ref{generalc1st})
under variation of the function $M(\tilde{x},t).$ Furthermore, one
has
$$(r\sin\phi)' = \tau_i\frac{\partial (r\sin\phi)}{\partial x_i}
= \tau_i\beta_i \chi(x - \tilde{x})\,,$$ where $\tau_i =
\varepsilon_{ik}n_k$ is the unit vector tangential to the flame
front. Taking into account also that $$\varepsilon_{ik}\beta_i =
\frac{v_{k+}}{v_+}\chi(x - \tilde{x})\,,$$ we find
$$(r\sin\phi)' = \varepsilon_{ik}n_k\beta_i\chi(x - \tilde{x}) =
\frac{v_{k+}n_k}{v_+} \equiv \frac{v^n_{+}}{v_+}\,.$$ Similarly,
$$(r\cos\phi)' = \frac{v_{k+}\tau_k}{v_+}
\equiv\frac{v^{\tau}_{+}}{v_+}\,.$$ Putting all these results
together and omitting the factor $e^{\nu t},$ we obtain the
following equation for the $x$-dependent parts of the perturbations
\begin{eqnarray}\label{generalc1lin}&&
2\tilde{\omega}' + \int\limits_{-\infty}^{+\infty}d\tilde{x}
\tilde{\xi}_{\alpha}(\tilde{x})\frac{\delta}{\delta
\xi_{\alpha}(\tilde{x})}\left(1 +
i\hat{\EuScript{H}}\right)\left\{[\omega]' -
\frac{Nv^n_+\sigma_+\omega_+}{v^2_+} \right\} \nonumber \\&& =
\frac{i\nu}{2}\left(1 + i\hat{\EuScript{H}}\right)
\int\limits_{-\infty}^{+\infty}d\tilde{x}\hat{M}_{\alpha}({\tilde{x}})
\tilde{\xi}_{\alpha}(\tilde{x})\frac{(v^n_{+}+iv^{\tau}_+)\omega_+}{v^4_+}
\exp\left(- \frac{\nu r}{v_+}e^{ - i\phi}\right)\chi(x -
\tilde{x})\,,
\end{eqnarray}
\noindent where $\delta/\delta \xi_{\alpha}(\tilde{x})$ denotes the
functional differentiation, and $\tilde{\omega} = \tilde{u} +
i\tilde{w}\,.$ Together with the linearized evolution equation,
Eq.~(\ref{generalc1lin}) can be used to carry out stability analysis
of general steady flame configurations, given by the solutions of
Eqs.~(\ref{evolutiongen}), (\ref{generalc1st}). To the best of our
knowledge, no such closed equation as (\ref{generalc1lin}) has yet
been derived to handle the problem. Its applications will be
presented elsewhere.

\section{Discussion and Conclusions}

The results presented in this paper solve the problem of
non-perturbative {\it description} of unsteady premixed flame
propagation with arbitrary gas expansion. Supplemented by the
evolution equation, Eq.~(\ref{generalc1}) gives the closed
description of unsteady flames in the most general form. Thus, as in
the stationary $2D$ case, the dilemma mentioned in Introduction is
resolved in negative for $2D$ unsteady flame propagation. The
conclusion that the detailed bulk structure of the gas flow is
actually unnecessary for describing front dynamics is even more
striking in the latter case. Indeed, the highly complicated vortex
flow downstream continuously changing in time is naturally expected
to have an exceedingly complicated nonlocal influence on the flame
front evolution. We proved, however, that all necessary information
about this influence is encoded in the complex history of the
combination $M = N\bar{v}^n_+\sigma_+\,.$ In this connection, a
curious circumstance is worth mentioning. As we saw in
Sec.~\ref{structure}, the vortex component depends on spatial
coordinates through the complex combinations $\tau_{\pm}$ appearing
as the limits of integration in the complex time plane. It is easy
to see that for any curved flame configuration, there are always
points at the front, where $\Omega <0$. For such points, the time
argument of the function $M(\tilde{x},t - \tau)$ in the integrand of
Eq.~(\ref{vortdef}) has real part $>t.$ In other words, integration
over such points is in a sense looking into the flame future. This
does not lead to any conflict with causality, because the
corresponding contribution is eventually annihilated by the operator
$(1 + i \hat{\EuScript{H}})$ in Eq.~(\ref{generalc1}). Indeed, the
vortical part of $\bm{v}^v$ comes from integration over $\tilde{x}$
such that the vectors $\bm{r}$ and $\bar{\bm{v}}_+$ are almost
parallel, i.e. $\Omega \approx 1,$ and therefore, the contribution
of points with $\Omega < 0$ is a pure potential satisfying
Eq.~(\ref{ch}). Retaining this contribution in intermediate formulas
is necessary to guarantee continuity of the potential component.

The paradox as to the seemingly ``teleological'' structure of
(\ref{generalc1}) is closely related to the analyticity properties
of $M(x,t)$ in the complex $t$-plane, that allowed us to simplify
Eq.~(\ref{kernel1}) to Eq.~(\ref{kernel3}). Our considerations were
carried under the very weak assumption of exponential boundedness of
this function, which is certainly sufficient for investigation of
any flame propagation phenomena. In particular, the linear stability
problem analyzed in Sec.~\ref{ldinstab} gives an example of $M(x,t)$
which is analytic in the complex plane, so that the question whether
the contour deformation is legitimate does not arise at all. In this
simplest case knowing the exponentials $\exp(\nu t)$ in
Eq.~(\ref{eq:5.3}) at current time $t$ allows one to predict their
future, so that the ``teleological'' question does not arise either.
Things are more interesting, however, in the case of flame
propagation in time-dependent gravitational field. Suppose that the
experimentalist plans to leave the burner fall freely at some time
instant $t_0.$ This means that $g(t),$ and hence, $M(\tilde{x},t)$
[Cf. Eqs.~(\ref{mfunction}), (\ref{gvort})] will have singularities
at $\tau_0 = t_0 \pm i\Delta t,$ where $\Delta t$ is of the order of
duration of the field switching off. As is shown in the Appendix,
crossing these singularities by the contour $C_- \cup C_+$ in Fig. 2
gives rise to a potential contribution that satisfies the conditions
a)--c) of Sec.~\ref{structure}, and hence does not change
Eq.~(\ref{generalc1}), thus resolving the causality issue.

To conclude, Eq.~(\ref{generalc1}) opens a wealth of key
developments in theoretical and numerical combustion (extended
propagation laws, coupling with acoustics, burners, etc.), not to
mention the other fronts evoked at the beginning of this paper. In
particular, Eq.~(\ref{generalc1lin}) allows direct analytical
investigation of small disturbances propagating on steady front
patterns such as Bunsen- or V-flames in 2-D configurations.
Extension of the above results to the three-dimensional problems is
still an open question, one of the main difficulties being the
generalization of the $(1+i \hat{\EuScript{H}})$ operator projecting
out the potential contributions of the burnt-gas flow.

\acknowledgements{The work presented in this paper was carried out
at the {\it Laboratoire de Combustion et de D\'etonique}. One of the
authors (K.A.K.) thanks the {\it Centre National de la Recherche
Scientifique} for supporting his stay at the Laboratory as a {\it
Chercheur Associ\'e}.}

\begin{appendix}

\section{}

When deriving the expression (\ref{vortdef}) for the vortex mode in
Sec.~\ref{structure} we have omitted the contribution of the contour
integral in Eq.~(\ref{kernel2}), retaining only the singular
contribution of the poles $\tau_{\pm}.$ That this operation respects
the property c) was already shown in Sec.~\ref{vortdefinition}. We
will now prove that Eq.~(\ref{vortdef}) does reproduce correctly the
near-the-front distribution of vorticity of the burnt gas flow. In
essence, the subsequent calculation reproduces the consistency check
given in the appendix A of \citep{kazakov2}. First of all, taking
into account the formula
\begin{eqnarray}\label{tauderive}
\frac{\partial\tau_{\pm}}{\partial x_i} = \pm \frac{i}{\bar{v}_+}
\left(\beta_i \mp i\frac{\bar{v}_{i+}}{\bar{v}_+}\right)
\end{eqnarray}
which is readily verified using the definitions (\ref{spoints}),
(\ref{vectorbeta}), one has
\begin{eqnarray*}
\varepsilon_{ki}\partial_k v^v_i & = &
\frac{1}{4}\frac{\partial}{\partial x_i} \left\{
\int\limits_{-\infty}^{+\infty}d\tilde{x}\frac{e^{-\mu
r}}{\bar{v}_+} \left[M(\tilde{x},t-\tau_+)\left(\beta_i -
i\frac{\bar{v}_{i+}}{\bar{v}_+}\right)\right.\right.\\
&& \hspace{3.2cm}\left.\left. + M(\tilde{x},t-\tau_-)\left(\beta_i +
i\frac{\bar{v}_{i+}}{\bar{v}_+}\right)\right]\rule{0pt}{20pt}\right\}_{\mu
= 0^+}
\end{eqnarray*}
\begin{eqnarray*}
&=& \frac{1}{2}\frac{\partial}{\partial x_i} {\rm Re}\left\{
\int\limits_{-\infty}^{+\infty}d\tilde{x}\frac{e^{-\mu
r}}{\bar{v}_+} M(\tilde{x},t-\tau_+)\left(\beta_i -
i\frac{\bar{v}_{i+}}{\bar{v}_+}\right) \right\}_{\mu = 0^+}\,.
\end{eqnarray*}
\noindent Following the argument given in Sec.~\ref{structure}, one
performs the differentiation with respect to $x_i$ under the sign of
the $\tilde{x}$-integral, and sees that the derivative of $e^{-\mu
r}$ leads to an integral proportional to $\mu .$ Therefore, the only
non-zero term contributed by this integral after $\mu $ is continued
to zero is an inessential $\bm{x}$-independent constant that falls
off from the expression ${\rm rot} D\bm{v}^v.$ Next, it is easily
checked that
\begin{eqnarray}\label{tauder2}
\left(\beta_k - i\frac{\bar{v}_{k+}}{\bar{v}_+}\right)^2 \equiv 0\,,
\end{eqnarray}
\noindent so differentiation of $M(\tilde{x},t-\tau_+)$ gives zero,
too. Note also that $\bm{\beta}$ can be written as
\begin{eqnarray}\label{betaeq}
\beta_i = \frac{\varepsilon_{ik}\bar{v}_{k+}}{\bar{v}_{+}}
\chi(\varepsilon_{lm}r_l \bar{v}_{m+})\,,
\end{eqnarray}
\noindent since it is orthogonal to $\bar{\bm{v}}_+$ and changes
sign at the point satisfying Eq.~(\ref{point1}). Thus, we find
\begin{eqnarray}
\varepsilon_{ki}\partial_k v^v_i &=& {\rm Re}\left\{
\int\limits_{-\infty}^{+\infty}d\tilde{x}\frac{e^{-\mu
r}}{\bar{v}^2_+} M(\tilde{x},t-\tau_+)
\varepsilon_{ik}\bar{v}_{k+}\varepsilon_{in}
\bar{v}_{n+}\delta(\varepsilon_{lm}r_l \bar{v}_{m+}) \right\}_{\mu =
0^+} \nonumber \\ &=& \int\limits_{-\infty}^{+\infty}d\tilde{x}
M(\tilde{x},t-r/\bar{v}_+) \delta(\varepsilon_{lm}r_l \bar{v}_{m+})
\,.\label{a1}
\end{eqnarray}
\noindent The factor $e^{-\mu r}$ and the symbol of analytic
continuation have been omitted in the last expression because it is
explicitly finite. The argument of the $\delta$-function turns into
zero when the vectors $r_i$ and $\bar{v}_{i+}$ are parallel. Near
this point, one has approximately
$$\varepsilon_{lm}r_l \bar{v}_{m+} = r \bar{v}_{+}\phi\,.$$

On the other hand, a simple geometric consideration shows that
$d\tilde{x}$ near the same point can be written as (see
Fig.~\ref{fig3})
$$d\tilde{x} = - \frac{r\bar{v}_{+} d\phi}{N\bar{v}^n_+}\,,$$ where
all quantities are taken at the time instant $t.$ Substituting these
expressions into Eq.~(\ref{a1}), and taking into account the
relation
$$\delta(\alpha x) = \frac{1}{|\alpha|}\delta(x)$$ yields
\begin{eqnarray}
\varepsilon_{ki}\partial_k v^v_i &=& \int d\phi\delta(\phi)
\frac{M(\tilde{x},t-r/\bar{v}_+)}
{N(\tilde{x},t)\bar{v}^n_+(\tilde{x},t)}\,.
\end{eqnarray}
\noindent If the observation point is taken at the flame front, $r$
turns into zero together with $\phi.$ Recalling also the definition
(\ref{mfunction}) of the function $M(\tilde{x},t),$ we arrive
finally at the desired identity
$$\left(\varepsilon_{ki}\partial_k v^v_i\right)_+ =
\sigma_+(\tilde{x},t)\,.$$

\begin{figure}
\centering
\includegraphics[width=.5\textwidth]{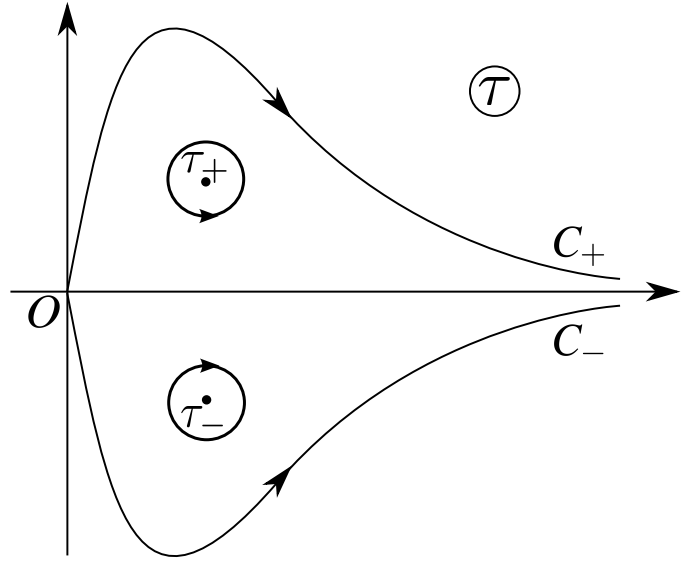}
\caption{Extraction of the singularity in Eq.~(\ref{kernel1}) by
contour deformation in the complex $\tau$-plane.}\label{fig2}
\end{figure}

\begin{figure}
\centering
\includegraphics[width=.5\textwidth]{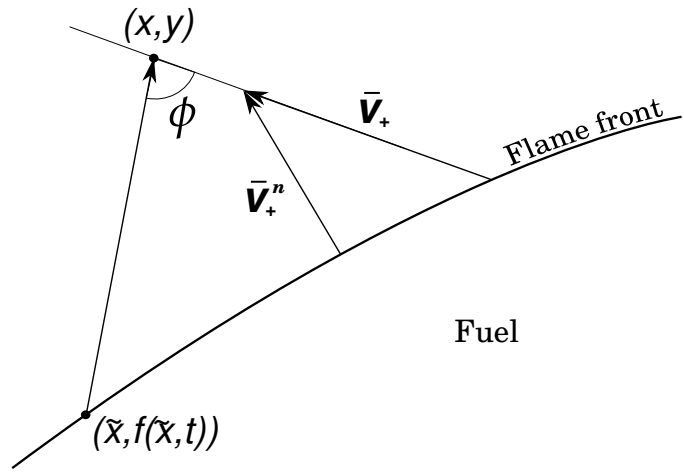}
\caption{Near-the-front structure of the flow downstream.}
\label{fig3}
\end{figure}

Let us now return to the question of possibility to perform the
contour deformation in the complex $\tau$-plane, used in the
derivation of Eq.~(\ref{vortexp}). We note, first of all, that this
question is concerned entirely with the structure of the potential
component of the gas velocity. Indeed, under our general assumption
of existence of a short wavelength cutoff, all functions involved
are smooth functions of time (because any structure of finite size
takes finite time to develop). Hence, these functions (in
particular, the function $M(\tilde{x},t)$) are analytic in a
vicinity of the real axis in the complex $t$-plane. On the other
hand, we know that the value of vorticity in any given point $(x,y)$
near the front is equal to its value at the point on the front,
satisfying $\Omega = 1,$ in which case the poles $\tau_{\pm}$ take
the real value $r/\bar{v}_+.$ For $\tilde{x}$ in a vicinity of that
point, $\tau_{\pm}$ belong to the analyticity domain of
$M(\tilde{x},t),$ and hence, of the function
$\EuScript{M}(\tilde{x},\tau,t).$ Thus, the use of the Cauchy
theorem and the contour deformation performed in
Sec.~\ref{structure} are legitimate for these $\tilde{x},$ while
integration over all other $\tilde{x}$ gives rise to a potential
contribution. This, however, does not conclude consideration,
because the argument just given proves potentiality of this
contribution in some vicinity of the given observation point. To
prove it for all $x \in [0,1],$ the contour $C_- \cup C_+$ in
Fig.~\ref{fig2} should be moved to the left of $\tau_{\pm}$ for all
$\tilde{x}.$ Let us show that this is possible indeed under the
assumption used already in the derivation of Eq.~(\ref{vortdef}),
namely, that $M(\tilde{x},t)$ considered as the function of the
complex $t$ is exponentially bounded near $t=\infty.$ Suppose, for
instance, that this function is meromorphic, i.e., has only poles of
arbitrary order in the $t$-plane. Any pole of the order $n$ in the
function $M(\tilde{x},t)$ becomes an $(n-1)$th order pole with
respect to $\tau$ in $\EuScript{M}(\tilde{x},\tau,t).$ Hence,
crossing these poles by $C_- \cup C_+$ does not change the right
hand side of Eq.~(\ref{kernel2}) unless $n=2$ or $n=1.$ Consider
first the case $n=2.$ Then $\EuScript{M}(\tilde{x},\tau,t)$ has
simple poles at some point $\tau_0$ and its complex conjugate
$\tau^*_0$ ($\tau_0$ may depend on $\tilde{x},t,$ but we do not
write this dependence explicitly, for brevity). We are to show that
their contribution to the right hand side of Eq.~(\ref{kernel2}),
given by the Cauchy theorem as (``c.c.'' stands for complex
conjugate)
\begin{eqnarray}\label{addn2}
\frac{2\pi i}{4\pi}\,{\rm
res}\,\EuScript{M}(\tilde{x},\tau_0,t)\left\{\frac{1}{\tau_0 -
\tau_+ } + \frac{1}{\tau_0 - \tau_- }\right\} + {\rm c.c.}\,,
\end{eqnarray}
\noindent gives rise to a field $V_i$ that satisfies conditions a) -
c). Substituting this expression into Eq.~(\ref{vint2v}) yields (we
do not introduce intermediate regularization because the
$\tilde{x}$-integral will be shown to converge)
\begin{eqnarray}\label{vfield}
V_i = - \frac{i}{4}\varepsilon_{ik}\int\limits_{-A}^{+A}d\tilde{x}
\,{\rm
res}\,\EuScript{M}(\tilde{x},\tau_0,t)\partial_k\left\{\frac{1}{\tau_0
- \tau_+ } + \frac{1}{\tau_0 - \tau_- }\right\} + {\rm c.c.}
\end{eqnarray}
\noindent The property a) is evidently satisfied. To prove b) we
write, using Eqs.~(\ref{tauderive}), (\ref{tauder2}) and
(\ref{betaeq})
\begin{eqnarray*}&&
\partial_k\int\limits_{-A}^{+A}d\tilde{x} \,{\rm
res}\,\EuScript{M}(\tilde{x},\tau_0,t)\partial_k\left\{\frac{1}{\tau_0
- \tau_+ } + \frac{1}{\tau_0 - \tau_- }\right\} \nonumber\\&& =
\int\limits_{-A}^{+A}d\tilde{x} \,{\rm
res}\,\EuScript{M}(\tilde{x},\tau_0,t)\left\{\frac{\partial^2_{k}\tau_+}{(\tau_0
- \tau_+)^2 } + \frac{\partial^2_{k}\tau_-}{(\tau_0 - \tau_-)^2
}\right\} \nonumber\\&& = 2i\int\limits_{-A}^{+A}d\tilde{x} \,{\rm
res}\,\EuScript{M}(\tilde{x},\tau_0,t)\delta(\varepsilon_{lm}r_l
\bar{v}_{m+})\left\{\frac{1}{(\tau_0 - \tau_+)^2 } -
\frac{1}{(\tau_0 - \tau_-)^2 }\right\} = 0\,,
\end{eqnarray*}
\noindent since $\tau_+ = \tau_-$ when the argument of the delta
function is zero. Thus, ${\rm rot}\bm{V} = 0\,.$ Last,
$\EuScript{M}(\tilde{x},\tau,t)$ is periodic in $\tilde{x},$ and
therefore, so is its pole. Taking into account also that $\tau_{\pm}
= O(|\tilde{x}|),$ $\partial_i\tau_{\pm} = O(1)$ for $|\tilde{x}|\to
\infty,$ one sees that the $\tilde{x}$-integral in
Eq.~(\ref{vfield}) is convergent in the limit $A\to \infty$ for all
$\bm{x}.$

In the case $n=1$ the function $\EuScript{M}(\tilde{x},\tau,t)$
contains a logarithmic singularity of the form
$a(\tilde{x},t)\ln(\tau - \tau_0)$ (and its complex conjugate).
Crossing this singularity leads to the $2\pi$ jump in
$\arg(\ln(\cdot))$ for all points of the contour $C_- \cup C_+,$
located at one side of the point $\tau_0.$ Hence, expression
(\ref{addn2}) is replaced in this case by the following
\begin{eqnarray*}
\frac{2\pi i a(\tilde{x},t)}{4\pi}\int\limits_{0}^{\tau_0} d\tau
\left\{\frac{1}{\tau - \tau_+ } + \frac{1}{\tau - \tau_- }\right\} +
{\rm c.c.}
\end{eqnarray*}
\noindent The proof of the properties a)--c) is exactly the same as
before.

Let us finally consider the case when the function $M(\tilde{x},t)$
has branch singularities. If these singularities are connected by a
number of cuts so that $M(\tilde{x},t)$ is meromorphic in the cut
$\tau$-plane, then so is the function
$\EuScript{M}(\tilde{x},\tau,t),$ and moving the contour of
integration beyond a cut results in a contribution to the right hand
side of Eq.~(\ref{kernel2}) of the form
$$\frac{1}{4\pi}\int\limits_{C_0}d\tau
[\EuScript{M}](\tilde{x},\tau,t)\left\{\frac{1}{\tau - \tau_+ } +
\frac{1}{\tau - \tau_- }\right\} + {\rm c.c.}\,,$$ where
$[\EuScript{M}](\tilde{x},\tau,t)$ denotes the jump\footnote{If
$\EuScript{M}(\tilde{x},\tau,t)$ is singular at $C_0,$ the above
integral can be replaced by the integral of this function over a
closed contour embracing the cut $C_0.$} of the function
$\EuScript{M}(\tilde{x},\tau,t)$ across the cut $C_0.$ If this cut
has finite length, then the above considerations again apply
literally. However, in the case of an infinite cut, the
$\tau$-integral is apparently divergent. This means that such cuts,
if any, are only allowed in regions where
$\EuScript{M}(\tilde{x},\tau,t)$ satisfies more restrictive
conditions than the exponential boundedness. We do not pursue
details here, because physical significance of such cuts is not
clear.

Thus, the function $M(\tilde{x},t)$ is allowed to have any number of
branch singularities in the complex $t$-plane, connected by cuts of
finite length, as well as any number of poles of arbitrary order to
justify the contour deformation used in Sec.~\ref{structure}.

\end{appendix}

\bibliography{references}
~\newpage
~\\\\
{\large\bf List of figures}\\\\
Elementary decomposition of the flow downstream used in the derivation of the expression (\ref{vint1})\dotfill 34\\
Extraction of the singularity in Eq.~(\ref{kernel1}) by contour deformation in the complex $\tau$-plane~~\dotfill 35\\
Near-the-front structure of the flow downstream\dotfill 36 ~\newpage
~\newpage
~\newpage

\end{document}